\documentclass[lettersize,journal]{IEEEtran}
\usepackage{amsmath,amsfonts}
\usepackage{algorithmicx}
\usepackage{algorithm}
\usepackage{array}
\usepackage{textcomp}
\usepackage{stfloats}
\usepackage{url}
\usepackage{verbatim}
\usepackage{graphicx}
\usepackage{cite}
\usepackage{svg}
\usepackage{booktabs, threeparttable, caption, makecell, subcaption}
\usepackage{multirow}
\usepackage{adjustbox}
\usepackage{algpseudocode}

\begin{document}

\title{Towards Lightweight Speaker Verification via Adaptive Neural Network Quantization}

\author{Bei Liu, 
        Haoyu Wang, \IEEEmembership{Student Member, IEEE,}
        and Yanmin Qian, \IEEEmembership{Senior Member, IEEE}
\thanks{
All the authors are with Department of Computer Science and Engineering \& MoE Key Laboratory of Artificial Intelligence, AI Institute, Shanghai Jiao Tong University, Shanghai, 200240 P. R. China (e-mail:\{beiliu, fayuge, yanminqian\}@sjtu.edu.cn)}

}

\markboth{Journal of \LaTeX\ Class Files,~Vol.~14, No.~8, August~2021}%
{Shell \MakeLowercase{\textit{et al.}}: A Sample Article Using IEEEtran.cls for IEEE Journals}


\maketitle

\begin{abstract}
Modern speaker verification (SV) systems utilize deep neural networks (DNN) to extract speaker embeddings. Nevertheless, these systems typically demand expensive storage and computing resources, thereby hindering their deployment on mobile devices. In this paper, we explore adaptive neural network quantization for lightweight speaker verification. Firstly, we propose a novel adaptive uniform precision quantization method which enables the dynamic generation of quantization centroids customized for each network layer based on k-means clustering. By applying it to the pre-trained SV systems, we obtain a series of quantized variants with different bit widths. To enhance the performance of low-bit quantized models, a mixed precision quantization algorithm along with a multi-stage fine-tuning (MSFT) strategy is further introduced. Unlike uniform precision quantization, mixed precision approach allows for the assignment of varying bit widths to different network layers. When bit combination is determined, MSFT is employed to progressively quantize and fine-tune network in a specific order. Finally, we design two distinct binary quantization schemes to mitigate performance degradation of 1-bit quantized models: the static and adaptive quantizers. Experiments on VoxCeleb  demonstrate that lossless 4-bit uniform precision quantization is achieved on both ResNets and DF-ResNets, yielding a promising compression ratio of $\sim$8. Moreover, compared to uniform precision approach, mixed precision quantization not only obtains additional performance improvements with a similar model size but also offers the flexibility to generate bit combination for any desirable model size. In addition, our suggested 1-bit quantization schemes remarkably boost the performance of binarized models. Finally, a thorough comparison with existing lightweight SV systems reveals that our proposed models outperform all previous methods by a large margin across various model size ranges.
\end{abstract}


\section{Introduction}
\IEEEPARstart{S}{peaker} verification (SV) involves the process of authenticating an individual's identity by analyzing the unique biometric traits embedded in voice. An automatic SV system can discern whether an enroll-test utterance pair originates from the same speaker or not. Typically, there exist two parts in a SV system: an embedding extractor that extracts speaker embeddings from utterances, and a similarity evaluator that assesses the similarity among the extracted embeddings. Traditionally, the prevalent approach for this task is the combination of i-vector~\cite{ivector} and probabilistic linear discriminant analysis (PLDA)~\cite{plda}. In recent years, the increasing prominence of deep learning has led to a widespread application of neural networks in this field, providing impressive results~\cite{tandem}. DNN-based SV systems generally comprises three main components: a feature extractor operating at frame-level, an embedding aggregator functioning at segment-level, and a speaker classifier. The process begins with a neural network generating a frame-level feature representation for a given utterance. Next, a temporal pooling layer is used to derive a fixed-length speaker embedding. Lastly, the whole system is trained using a multi-class speaker classifier. To enhance systems' performance, researchers have exerted numerous endeavors in multiple aspects, such as network backbones~\cite{tdnn-sv, xvector, ext-xvector, rvector, d-tdnn, ecapa, ecapa++, dense-residual, adaptive-cnn, df-resnet, df-resnet-journal, mlp}, pooling strategies~\cite{pooling1, pooling2, pooling3, pooling4, pooling5}, and training criteria~\cite{sv-loss1, sv-loss2, sv-loss3, sv-loss4}.

Regarding network backbones, various architectures have emerged over the past few decades. ~\cite{tdnn-sv} introduces the initial application of Time delay neural network (TDNN) in speaker verification as a replacement for the traditional i-vector. Following that, different variants, including x-vector~\cite{xvector} and E-TDNN~\cite{ext-xvector}, are designed to improve performance. Afterwards, an extraordinary milestone model called ECAPA-TDNN~\cite{ecapa} is proposed which incorporates multiple architectural enhancements to x-vector, leading to exceptional results. Recently, the newly developed ECAPA++~\cite{ecapa++} has attained the state-of-the-art performance by focusing on fine-grained speaker information. Aside from TDNN-based systems, the winner of VoxSRC-2019~\cite{rvector} showcases the surprising success of utilizing 2D convolutional neural network (CNN) for the SV task. Since then, ResNet~\cite{ResNet} has gained widespread popularity in this field. To further boost representation capability, several lightweight attention modules have been devised that can be easily integrated with ResNet~\cite{duality-att, dpnet, simple-att, aff}. On the other hand, recent studies have unveiled that increasing the depth of neural networks can lead to consistent improvements in performance. For example, ~\cite{df-resnet, df-resnet-journal} introduce a depth-first version of ResNet, significantly boosting the network's depth to an impressive 233 layers. Additionally, \cite{cnsrc-2022} reaches a new level by extending ResNet's layer number to 293. 
   
Although larger and more advanced architectures have led to significant performance improvements, the extensive storage and computational demands of these systems generally pose obstacles to deploying them on mobile devices. In fact, designing lightweight speaker verification systems tailored for mobile devices is an urgent and challenging task. In prior works, different approaches have been explored for small-footprint speaker verification, including knowledge distillation~\cite{kd-shuai, skd}, network quantization~\cite{binary-lm, quant-ljy} and efficient architecture designs~\cite{julien, ecapa-lite, cs-ctcsconv1d}. Knowledge distillation~\cite{kd} is a widely employed technique for model compression, which involves the transfer of knowledge from teacher networks to student counterparts. Despite the potential to boost students' performance without increasing model size, deploying these networks on mobile devices still remains cumbersome due to considerable parameters. Besides, network quantization~\cite{deep-compress} offers a promising way to decrease model size by utilizing a reduced bit width to represent full-precision weights. However, the existing quantization methods suffer from a significant performance gap between quantized models and their full-precision counterparts, particularly in low-bit scenarios. In addition, many researchers are dedicated to intricately crafting efficient operators and lightweight network architectures. Although there is a notable reduction in parameter number and computational complexity, it can incur severe performance degradation, barely fulfilling the demands of real-life SV applications. This paper delves into the exploration of adaptive neural network quantization with the aim of achieving a better trade-off on performance and model size in the context of lightweight speaker verification. Specifically, three different types of adaptive strategies are investigated, including adaptive uniform precision quantization, mixed precision quantization and adaptive binary quantization.

Firstly, we propose a novel adaptive uniform precision quantization method. Previous studies~\cite{binary-lm, quant-ljy} have made attempts to employ quantization techniques in SV systems. Nevertheless, quantization levels utilized in existing methods are manually crafted, and remain fixed for each network layer. For example, the quantized values in~\cite{binary-lm} are limited to $\left\{-1, +1\right\}$ in the entire network. Similarly, both Uniform and Power-of-Two (PoT) quantization in~\cite{quant-ljy} adopt identical quantization levels across all network layers, which are pre-defined based on empirical observations. In fact, evidences from~\cite{apot} have revealed that the distribution of full-precision weights within a neural network exhibits significant variations across different layers. The direct adoption of pre-determined and fixed quantization levels for all layers will incur substantial quantization errors, especially in low-bit cases. In contrast to previous methods, we present a novel adaptive uniform precision quantization technique that allows for the dynamic generation of quantization centroids specific to each layer based on k-means clustering. By applying it to the pre-trained SV systems, we obtain a series of quantized variants with different bit widths.

Moreover, to improve the performance of quantized models with low bit width, we introduce a mixed precision quantization algorithm combined with a multi-stage fine-tuning (MSFT) strategy. In the context of uniform precision quantization, all network layers employ the same bit width. However, it ignores the crucial fact that each layer displays different sensitivities to quantization. To fully unleash the potential of network quantization, we present a mixed precision quantization approach that can effectively assign higher bit width to more sensitive layers and lower bit width to less sensitive ones. Specifically, the sensitivity is measured using second-order information, namely the Hessian eigenvalues of weights. To select bit combination for a target model size, candidates in the search space are sorted according to their sensitivity values. Afterwards, we implement a multi-stage fine-tuning (MSFT) strategy to rescue the performance by progressively quantizing and fine-tuning the network in a specific order.

Finally, we design two advanced binary quantization schemes specifically crafted for 1-bit scenario. Unlike other forms of quantization, 1-bit quantization is the extreme case where weights are mapped to merely two values. Previous works typically adopt integer set such as $\left\{-1, +1\right\}$~\cite{bnn} or $\left\{0, +1\right\}$~\cite{siman}. Despite a significant compression ratio, this will severely weaken the representation ability of networks, which inevitably leads to notable performance degradation. To reduce quantization errors in binarized models, we develop two distinct binary quantization schemes: the static and adaptive quantizers. The static scheme incorporates an entropy-preserving weight regularization technique to address the magnitude mismatch between real-valued and quantized weights. Meanwhile, the adaptive quantizer dynamically generates binary sets across different layers, ensuring a better alignment with the real-valued weight distribution. Both of them substantially boost the performance of binarized networks.

In summary, the main contributions of this work are enumerated as follows:

\begin{enumerate}
    \item Firstly, we present a novel adaptive uniform precision quantization approach based on k-means clustering, achieving lossless 4-bit compression for both ResNets and DF-ResNets.
    \item To enhance the performance of low-bit quantized models, a mixed precision quantization algorithm along with a multi-stage fine-tuning (MSFT) strategy is further introduced. Unlike uniform precision method, mixed precision quantization achieves better results with similar model sizes. Moreover, it offers the flexibility to generate bit combination for any desirable model size.
    \item Afterwards, two advanced quantization schemes are devised, which are tailored for 1-bit quantization. Both of them notably improve the representation capacity of binarized models.
    \item We conduct performance evaluations of the newly proposed quantization algorithms on several strong speaker verification models, including ResNet34, ResNet101, DF-ResNet110 and DF-ResNet179, demonstrating their generality and robustness. 
    \item Plus, a detailed visualization and analysis of the quantized weight distribution are provided, confirming the effectiveness of our methods. 
    \item Finally, a thorough comparison with existing lightweight SV systems illustrates that our resulting models significantly outperform previous ones across various model size ranges.
\end{enumerate}

This paper is an extension of our previous work~\cite{adaptive-quant}. In this article, we first introduce an enhanced version of k-means quantization by introducing two different rescaling strategies. Moreover, two advanced 1-bit quantization schemes are designed to improve the performance of binarized models. In the experiments, we include an extended evaluation of popular models like ResNet101, DF-ResNet110 and DF-ResNet179. In addition, a detailed weight distribution analysis and an extensive comparison with other SV systems are provided, illustrating the effectiveness and superiority of our approaches.

\section{Lightweight Speaker Verification}
Recently, lightweight speaker verification has emerged as a crucial and active research area. The existing methods can be summarized as follows.

\subsection{Knowledge Distillation}
 As a popular model compression technique, knowledge distillation has been extensively explored in the SV field~\cite{kd-shuai, skd, kd-zly}. ~\cite{kd-shuai} presents two different distillation strategies for small-footprint speaker embedding learning: label-level and embedding-level. ~\cite{kd-zly} attempts to improve single-modality system by transferring knowledge from pre-trained multi-modality one. ~\cite{skd} introduces a self-knowledge distillation framework where teacher and student models are jointly trained. 

\subsection{Network Quantization}
Network quantization is another effective means of shrinking model size and lowering memory expenses. ~\cite{binary-lm} achieves the successful binarization of training weights for SV systems, resulting in memory savings of up to 32x. ~\cite{quant-ljy} delves into the application of Uniform and PoT quantization to both ResNet and ECAPA-TDNN, attaining 8-bit compression without notable performance loss. In addition, information probing analysis demonstrates the quantized models' ability to retain the essential speaker-related knowledge.

\subsection{Efficient Architecture Designs}
Considerable endeavors have been made to devise efficient operators and lightweight architectures aimed at reducing model complexity. For example, ~\cite{julien} proposes a lightweight two-stage model based on 1D time-channel separable convolutional module. ~\cite{ecapa-lite} introduces a lite version of ECAPA-TDNN by reducing feature dimension and utilizing depth-wise separable convolution. ~\cite{cs-ctcsconv1d} presents an enhanced lightweight module, namely CTCSConv1D, for small-footprint speaker verification. Plus, Thin-ResNet~\cite{thin-resnet2} and Fast-ResNet~\cite{fast-resnet} are two light variations of ResNet.

\section{Adaptive Uniform Precision Quantization}
In this section, we first introduce the basic concepts of network quantization. Subsequently, we conduct an in-depth exploration of the proposed adaptive uniform precision quantization methods based on k-means clustering.

\subsection{Preliminaries}
Network quantization, a common approach for model compression, primarily involves two essential operations: quantization and dequantization.

\begin{figure}[!t]
  \centering
  \includegraphics[width=0.95\linewidth]{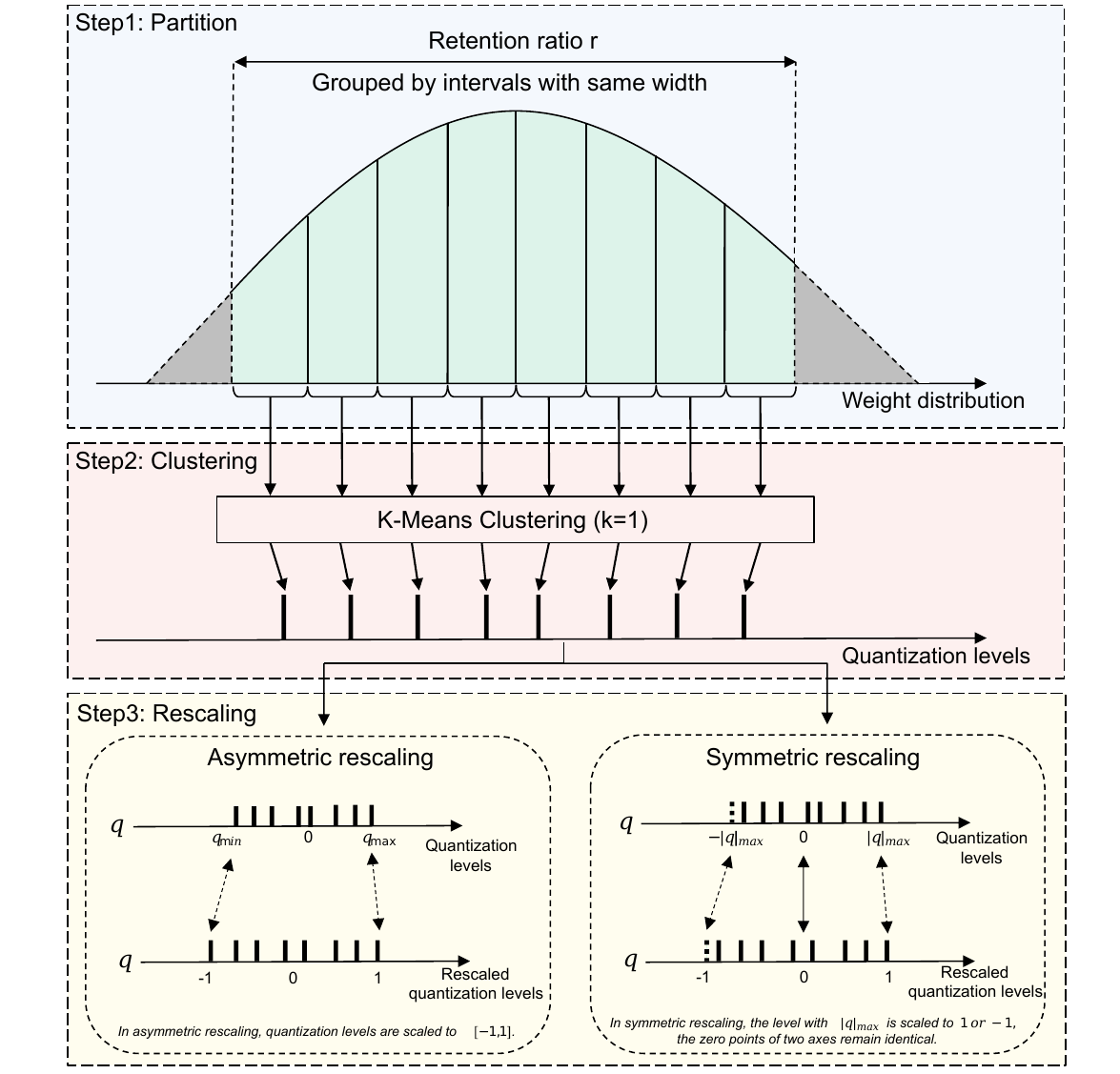}
  \caption{The pipeline of k-means clustering based quantization levels. Using 3-bit quantization as an illustration, this process involves three steps: partition, clustering and rescaling to yield eight quantization centroids.}
  \label{fig:k_means}
\end{figure}

\textbf{quantize operation: }The goal of this step is to map floating-point numbers to discrete values with lower precision. In previous works, quantization levels are pre-defined and remain fixed for each layer. For example, quantization centroid $q$ for $n$-bit Uniform~\cite{uniform}, Power-of-Two (PoT)~\cite{pot} and Additive Power-of-Two (APoT)~\cite{apot} quantization can be represented as follows:
\begin{equation}
  \text{Uniform}: q \in \{0, \frac{\pm1}{2^{n-1}-1}, \frac{\pm2}{2^{n-1}-1}, \ldots, \pm1 \}
  \label{eq1}
\end{equation}
\begin{equation}
  \text{PoT}: q \in \{0, \pm 2^{-2^{n-1}+1},\pm2^{-2^{n-1}+2}, \ldots, \pm 1 \}
  \label{eq2}
\end{equation}
\begin{equation}
\begin{split}
    \text{APoT}: q \in \left\{0,\pm({2^{-i}}+{2^{-j}}), \ldots, \pm 1 \right\}
\end{split}
\label{eq3}
\end{equation}
This process can be performed using \textit{round} function.

\textbf{dequantize operation: }This procedure involves an affine transformation from fixed-point values to full-precision numbers. The conversion formula is presented below:
\begin{equation}
  Q = \alpha \times q
  \label{eq4}
\end{equation}
where $\alpha$ denotes a learnable scaling factor with full precision. $Q$ is the quantized version of a given full-precision number. For a network, we can construct a set of quantization levels $q$ and its associated scaling factor $\alpha$ for every layer within it. 

There exist two different quantization frameworks: post-training quantization (PTQ) and quantization-aware training (QAT). Both frameworks include the quantize and dequantize operations.

\textbf{post-training quantization: }PTQ~\cite{ada_round, brecq} directly perform quantization to pre-trained weights without requiring any further training. Despite the simplicity, it can cause significant performance degradation.

\textbf{quantization-aware training: }On the contrary, QAT~\cite{dorefa, pact, fake-quant} introduces a fake node into the computational graph to simulate the quantization process. Through gradient approximation, system can be trained in an end-to-end manner, thereby recovering accuracy loss caused by quantization. Our work is based on the framework of QAT. Specifically, it is a technique that takes the effects of quantization on a model's performance into consideration. During the training process, a full-precision copy of the model weights is retained to mitigate quantization errors. In addition, straight-through estimation (STE)~\cite{ste} is adopted to directly assign incoming gradients as outgoing gradients for quantization operation (Eq. \ref{eq5}). Once the model is trained, only the quantized weights are utilized for inference.
\begin{equation}
  \frac{\partial \mathcal{L}}{\partial Q} = \frac{\partial \mathcal{L}}{\partial W}
  \label{eq5}
\end{equation}
where $\mathcal{L}$ means the loss function. $W$ represents full-precision weights of a network.

\begin{figure}[!t]
  \centering
  \includegraphics[width=0.95\linewidth]{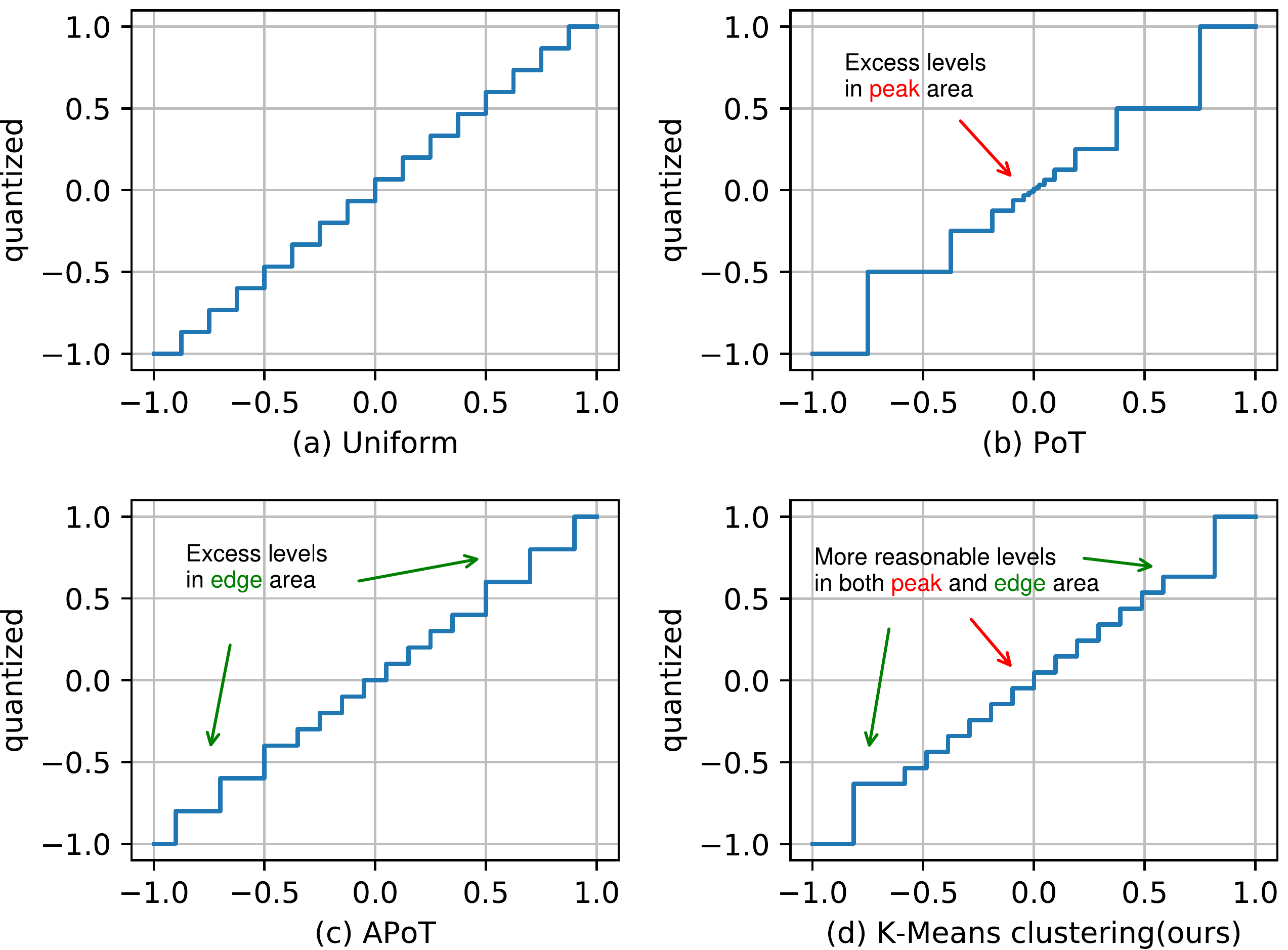}
  \caption{The comparison of different 4-bit quantization levels, including Uniform, PoT, APoT and k-means clustering (ours).}
  \label{fig:quant_levels}
\end{figure}

\subsection{Adaptive Uniform Precision Quantization}
\label{ssec:kmqat}
Conventionally, quantization levels have been pre-established using a heuristic approach. Uniform quantization~\cite{uniform}, as illustrated in Fig. \ref{fig:quant_levels}(a), employs ones that are evenly spaced. In fact, several studies~\cite{deep-compress, hawq, hawq-v2} have unveiled that network weights typically conform to a bell-shaped distribution. (Fig. \ref{fig:k_means}). Even though PoT and APoT introduce an improved version of quantization levels, a notable discrepancy between quantization centroids and weight distribution still exists. For example, excessive levels are allocated to the peak area of the weight distribution in PoT, whereas APoT places disproportionate emphasis on the edge area, as Fig. \ref{fig:quant_levels}(b) and (c) depict. Moreover, the existing methods utilize the same centroids for all layers within a network, resulting in substantial quantization errors due to variations in weight distribution across different layers. In this section, we present a novel adaptive uniform precision quantization technique, which enables the dynamic generation of quantization centroids customized for each network layer by applying k-means clustering to full-precision weights. Our proposed method consists of three steps. Fig. \ref{fig:k_means} schematically displays the pipeline of k-means clustering based quantization levels. The specific details are provided below.

\textbf{step 1: }This step aims to create a more reasonable division of the real-valued weight distribution. To mitigate the effect of \textit{outliers}, we firstly introduce the concept of retention ratio $r$, which represents the proportion of weights that are preserved. As shown in Fig. \ref{fig:k_means}, the tail part of weight distribution (grey area), accounting for a total of $1-r$, is discarded directly. Then, we evenly partition the weights of each layer into $2^n$ intervals for $n$-bit quantization, ensuring that each portion of the weight distribution will be allocated an appropriate centroid. In the ablation study, we explore different retention ratios: $\{ 100\%, 90\%, 80\% \}$, as outlined in Table \ref{table:1}. It is clear that $r=90\%$ yields the best results. Therefore, we decide to adopt a retention ratio of $90\%$ for all following experiments.

\begin{table}[t]
  \footnotesize
  \caption{The ablation study on retention ratio $r$ in partition.}
  \label{table:1}
  \centering
  \begin{adjustbox}{width=.48\textwidth,center}
  \begin{tabular}{ccccc@{\extracolsep{3pt}}}
    \toprule
    \textbf{Model}  & {\makecell{\textbf{Retention} \\ \textbf{Ratio $r$}} }&{\makecell{\textbf{Vox1-O} \\ \textbf{EER(\%)}} } &{\makecell{\textbf{Vox1-E} \\ \textbf{EER(\%)}}} &{\makecell{\textbf{Vox1-H} \\ \textbf{EER(\%)}} } \\
    \midrule
    \multirow{3}{*}{\makecell{ResNet34 \\ + 4-bit} }  &  100\% & \textbf{0.925} & 1.047 & 1.914  \\
    &  \textbf{90\%} &0.930 &\textbf{1.033}&\textbf{1.899} \\
    &  80\% &0.963&1.039&1.910 \\
    \bottomrule
  \end{tabular}
  \end{adjustbox}
\end{table}

\begin{table}[t]
  \footnotesize
  \caption{The ablation study on asymmetric rescaling and symmetric rescaling.}
  \label{table:2}
  \centering
  \begin{adjustbox}{width=.48\textwidth,center}
  \begin{tabular}{ccccc@{\extracolsep{3pt}}}
    \toprule
    \textbf{Model}  & {\makecell{\textbf{Rescaling} \\ \textbf{Mode}} }&{\makecell{\textbf{Vox1-O} \\ \textbf{EER(\%)}} } &{\makecell{\textbf{Vox1-E} \\ \textbf{EER(\%)}}} &{\makecell{\textbf{Vox1-H} \\ \textbf{EER(\%)}} } \\
    \midrule		
    \multirow{2}{*}{\makecell{ResNet34 \\ + 4-bit} }  & asymmetric & \textbf{0.899} & 1.049 & 1.928  \\
    &  \textbf{symmetric} &0.930&\textbf{1.033}&\textbf{1.899} \\

    \bottomrule
  \end{tabular}
  \end{adjustbox}
\end{table}

\textbf{step 2: }In this step, we independently perform k-means clustering on each partitioned interval. In general, k-means clustering algorithm~\cite{k-means} seeks to divide data points into $k$ clusters with the objective of minimize the total squared distance within each cluster. The mean of data points in each cluster is called centroid. For each interval derived from step 1, we individually conduct k-means clustering with $k=1$, and treat the resulting cluster centroid as a quantization level. This ensures the minimum disparity between quantization centroids and weight distribution. Notably, our method has the ability to adaptively generate distinct quantization levels for each network layer based on the real-valued weight distribution. This feature can lead to a substantial decrease in quantization errors.

\textbf{step 3: }We empirically observe that the majority of quantization centroid $q$ from step 2 fall within the magnitude range of $10^{-3}\sim10^{-2}$, potentially causing numerical instability in scaling factor $\alpha$ learning during quantization-aware training. As Fig. \ref{fig:k_means} displays, we propose two different affine transformations, namely asymmetric and symmetric rescaling, to project quantization centroids from step 2 into $[-1, 1]$ for training stability. Notably, this is the first work to apply an affine transformation to the quantization centroids.

In asymmetric rescaling, we perform a direct mapping of $q_{min}$ to $-1$ and $q_{max}$ to $1$. The formula is as follows:
\begin{equation}
 q_{rescale} =  \frac{q-q_{min}}{q_{max}-q_{min}} \times 2 - 1
  \label{eq6}
\end{equation}

In contrast, symmetric rescaling aligns $-q_{max}$ with $-1$ and $q_{max}$ with $1$, ensuring that zero points of two axes remain identical. The rescaling procedure is provided below.
\begin{equation}
 q_{rescale} =  \frac{q-q_{min}}{q_{max}} - 1
  \label{eq7}
\end{equation}

As Table \ref{table:2} indicates, symmetric rescaling achieves better performance than asymmetric counterpart. Thus, we will utilize symmetric rescaling for all future experiments.

Through the above three steps, our resulting quantization levels $q_{rescale}$ can better align with the weight distribution across all network layers (Fig. \ref{fig:quant_levels}(d)), thereby enjoying superior representation capability. Combined with quantization-aware training technique, our approach is named as \textbf{K}-\textbf{M}eans based \textbf{Q}uantization \textbf{A}ware \textbf{T}raining (KMQAT).

\begin{figure*}[th]
  \centering
  \includegraphics[width=0.95\linewidth]{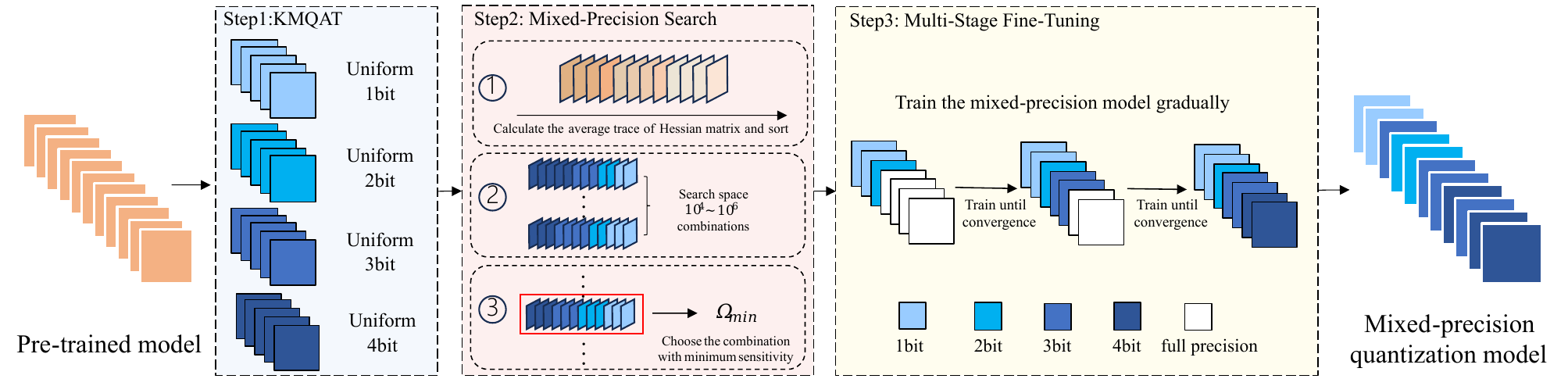}
  \caption{The pipeline of mixed-precision quantization. It consists of three steps. Firstly, KMQAT is utilized to produce uniform precision quantized models. Then, mixed-precision search is performed to generate the optimal bit combination for each network layer. Finally, we introduce multi-stage fine-tuning strategy to quantize and fine-tune network in a progressive manner.}
  \label{fig:mixed_precision}
\end{figure*}

\section{Mixed Precision Quantization}
\label{sec:mixed_precision}
The uniform precision approach described in previous section can achieve 4-bit quantization with a negligible reduction in performance. However, in low-bit scenarios, a noticeable performance gap still remains between quantized models and their full-precision counterparts. One potential explanation is that uniform precision quantization allocates the same bit width to each layer of a network, ignoring the aspect that different layers have varying sensitivities to quantization. This might result in a sub-optimal bit assignment. In this section, we introduce a mixed precision quantization algorithm aimed at boosting the performance of low-bit quantized models. It enjoys the advantage of dynamically assign diverse bit widths to each layer based on sensitivity analysis. Fig. \ref{fig:mixed_precision} illustrates the process of mixed precision quantization, and the detailed explanations are given below.

\textbf{step 1: }Given a pre-trained model and a target model size, the first thing is to identify a suitable set of candidate bit widths for each network layer. For low-bit cases, the set of candidates $\mathcal{C}$ is defined as $\{2, 3, 4 \}$ or $\{ 1, 2, 3, 4 \}$. Then, we utilize KMQAT proposed in Section \ref{ssec:kmqat} to produce uniform precision quantized models, as shown in Fig. \ref{fig:mixed_precision}. These quantized weights will then be included in the overall sensitivity calculation in the following step.

\textbf{step 2: }This step aims to find an optimal bit combination for each layer of a pre-trained network, guided by sensitivity analysis. The search process can be divided into three separate sub-steps. In contrast to uniform precision quantization, our mixed precision approach involves assigning varying bit precision to each layer based on its quantization sensitivity. Specifically, layers that are more sensitive to quantization will be allocated a higher bit width, while those less sensitive receive a lower one. Inspired by~\cite{hawq-v2}, the sensitivity of each layer is estimated using second-order information of the pre-trained weights, i.e. Hessian matrix. Further details are presented below.

\textit{Estimate Hessian Sensitivity: }As illustrated in Fig. \ref{fig:mixed_precision}, we initially compute the average trace of Hessian matrix for the pre-trained weights, employing this as the sensitivity metric for each layer. Subsequently, all network layers are re-arranged in a descending according to their average traces. Layers with a large average trace will be given a higher bit width, and conversely for those with a low average trace. This is reasonable because when layers exhibit a significant average trace, indicating a higher vulnerability to quantization impacts, it becomes necessary to allocate a greater bit precision to mitigate performance degradation.

\textit{Define Search Space: }Next, we need to establish the search space of bit combinations. For a candidate set $\{ 1, 2, 3, 4 \}$, each layer has four options, leading to a search space that grows exponentially with the number of network layers. This exponential expansion makes bit combination search infeasible. To shrink the search space, we simply partition the re-arranged network layers from previous step into four sections, as depicted in Fig. \ref{fig:mixed_precision}. Layers within the same section are assigned an identical bit width. As a result, the final search space $\mathcal{S}$ contains $10^{4}\sim10^{6}$ combinations, greatly speeding up the search process.

\textit{Determine Bit Combination: }This step focuses on selecting the optimal bit combination $s$ from earlier established search space $\mathcal{S}$. By incorporating second-order quantization perturbation, we define the total sensitivity of a quantized network for a specific bit combination $s$ as follows:
\begin{equation}
 \Omega_s = \sum_{i=1}^{L} \Omega_{i}=\sum_{i=1}^{L} \overline{\operatorname{Tr}}\left(H_i\right) \lVert W_i-Q_i \rVert_{2}^{2}
  \label{eq8}
\end{equation}
where $\Omega_{i}$ signifies the estimated sensitivity of the $i$-th layer, and $\Omega_{s}$ indicates the total sensitivity of a network with $L$ layers. $H_i$ represents Hessian matrix of the $i$-th layer. $\overline{\operatorname{Tr}}\left(H_i\right)$ denotes the average trace of $H_i$. $W_i$ is pre-trained weights in the $i$-th layer and $Q_i$ is the corresponding quantized weights. $\lVert W_i-Q_i \rVert_{2}^{2}$ is the second-order perturbation incurred by quantization on pre-trained weights.

The search process can be converted to an optimization problem in the following way:
\begin{equation}
  \begin{aligned}
  \min_{s} \quad & \Omega_s\\
  \textrm{s.t.} \quad & s \in \mathcal{S} \\
    & \text{model size} \leq \text{target model size} \\
  \end{aligned}
  \label{eq9}
\end{equation}

The entire mixed-precision search process can be efficiently completed on a CPU within a few minutes.

\begin{figure*}[!t]
  \centering
  \includegraphics[width=0.95\linewidth]{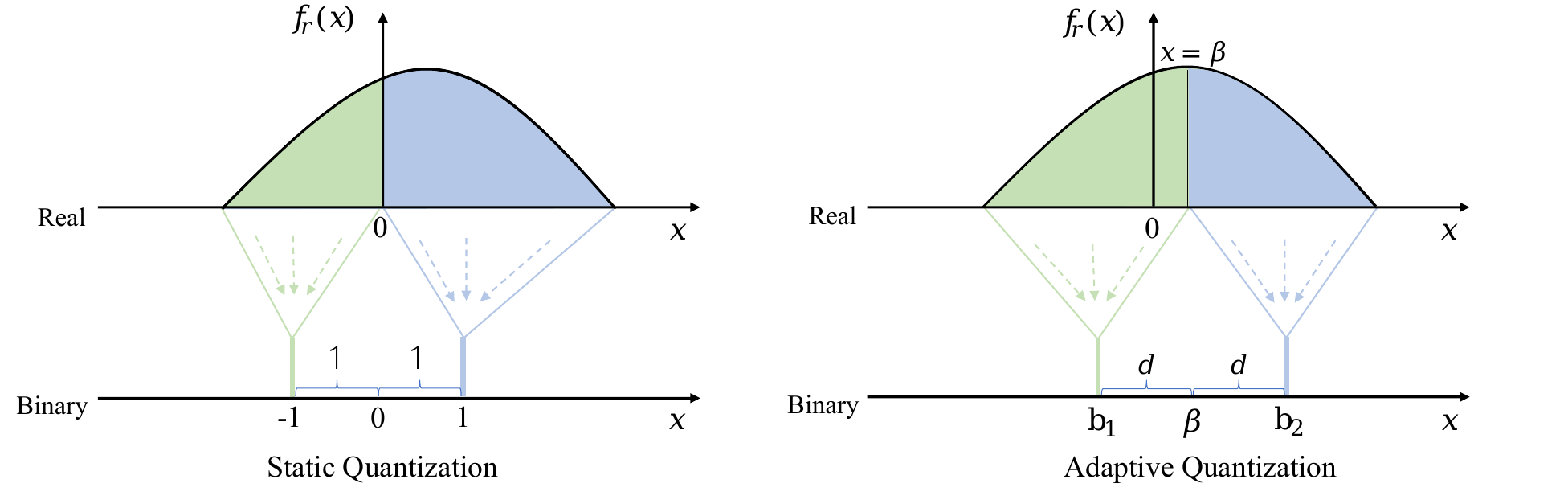}
  \caption{The overview of both static and adaptive binary quantization. Static quantization (left) projects real-valued weights into a fixed binary set $q \in\left\{-1, +1\right\}$ across all layers. In contrast, adaptive quantization (right) is capable of flexibly selecting a distinct binary set $Q \in\left\{b_1, b_2 \right\}$ for each layer to better align with the distribution of real-valued weights.}
  \label{fig:binary_quantizer}
\end{figure*}

\textbf{step 3: }After obtaining mixed-precision bit combination $s$, we will perform quantization aware fine-tuning in this step. In uniform precision quantization, the entire network undergoes a single round of quantization and fine-tuning. However, this approach might result in a sub-optimal solution for mixed precision quantization. Previous studies~\cite{inq, stoch_quant, hawq} have revealed that quantization errors can vary significantly across different network layers. Layers with a higher bit width typically exhibit smaller quantization errors, while those with a lower bit width can have much larger errors. Quantizing weights all at once in a single iteration can lead to inappropriate gradient direction during training, causing the model to converge to a sub-optimal local minimum. Instead, we propose a \textbf{M}ulti-\textbf{S}tage \textbf{F}ine-\textbf{T}uning (MSFT) strategy where only parts of network are quantized and fine-tuned at each stage. Specifically, MSFT progressively quantizes layers based on their bit widths. As Fig. \ref{fig:mixed_precision} shows, it starts with low bits and gradually progresses to high ones. Before going to the next stage, the partially quantized models will be fully trained until convergence. Algorithm \ref{alg::msft} provides a detailed description of MSFT.

\begin{algorithm}[h]
  \caption{Multi-stage fine-tuning (MSFT) for mixed-precision quantization}
  \label{alg::msft}
  \begin{algorithmic}[1]
    \Require
      $L$: the number of network layers;
      $f$: quantization function;
      $W$: pre-trained weights;
      $s$: bit combination obtained in step 2;
      $\mathcal{C}$: candidate set;
      $j$: $ 1 \leq j \leq \lvert \mathcal{C} \rvert$.
    \Ensure
      Mixed-precision quantized model
    \State Initial $W=\left \{W_1,W_2,...,W_L\right \}$; $s=\left \{s_{1}, s_{2}, ..., s_{L}\right \}$; $\mathcal{C}=\left \{c_{1},c_{2},...,c_{n}\right \}$ where $c_{1}<c_{2}<...<c_{n}$; $j=1$
    \Repeat
    \For{$i$ in $\left \{1,2,...,L\right \}$}
        \If{$s_{i} \le c_{j}$}
            \State $W_i \gets f(W_i)$;
        \Else 
            \State $W_i \gets W_i$;
        \EndIf
    \EndFor
    \State Train $W$ until convergence;
    \State $j \gets j+1$;
    \Until{The model is fully quantized}
  \end{algorithmic}
\end{algorithm}

\section{Extremely Low Bit Quantization}
Different from other bit quantizations, 1-bit quantization represents the most extreme case, projecting full-precision weights into merely two values. Despite the highest compression ratio, this method considerably impairs the network's representational capacity, inevitably causing significant performance decline. In this section, we introduce two distinct binary quantization schemes to reduce quantization errors in binarized models: the static and adaptive quantizers, both specifically devised for 1-bit quantization. Fig. \ref{fig:binary_quantizer} provides a comparative overview of both static and adaptive binary quantization. The in-depth discussion is presented below.

\subsection{Static Binary Quantization}
\label{ssec:static}
In traditional 1-bit quantization, a substantial performance disparity often exists between binarized models and full-precision ones. A possible reason could be the significant mismatch in magnitude between real-valued and quantized weights. For static 1-bit quantization, the binary values for all network layers are limited to a fixed set of integers, namely $\left\{-1, +1\right\}$, as Fig. \ref{fig:binary_quantizer} (left) exemplifies. However, empirical observations reveal that most weights fall within the $10^{-3}\sim10^{-2}$ range, as Section \ref{ssec:kmqat} points out. Such a mismatch in magnitude can lead to significant errors in quantization. To boost the performance of binarized network, we introduce an entropy-preserving weight regularization approach, inspired by~\cite{n2uq}, to alleviate this mismatch.

From an information theory perspective, a distribution possessing greater entropy is capable of retaining more information. Hence, we propose a weight regularizer designed to maintain maximum entropy and reduce the loss of information in quantized weights. Theoretical analysis reveals that the peak information entropy can be reached by equally partitioning real-valued weights across quantization levels. For binary integer set $\left\{-1, +1\right\}$, the distribution of quantized weights becomes approximately uniform when weights are regularized through below equation:
\begin{equation}
  W_i^{\prime} = \frac{\lvert W_i \rvert}{{\lVert W_i \rVert}_{l1}} W_i
  \label{eq10}
\end{equation}
where $W_i$ is real-valued weights of the $i$-th layer. $\lvert W_i \rvert$ represents the total count of elements in $W_i$. ${\lVert W_i \rVert}_{l1}$ refers to the L1 norm of $W_i$.

The resulting $W_i^{\prime}$ are then binarized using the following quantize and dequantize operations.
\begin{equation}
  q = round((clip(w^{\prime}, -1, 1)+1)\times \frac{1}{2}) \times 2 -1 \in\left\{-1,  1\right\}
  \label{eq11}
\end{equation}
\begin{equation}
  Q = \alpha \times q
  \label{eq12}
\end{equation}
where $w^{\prime}$ is an element of the matrix $W_i^{\prime}$. $clip$ function clamps weights between -1 and 1. $round$ is used to map values to the closest integer.

\subsection{Adaptive Binary Quantization}
\begin{figure}[!t]
  \centering
  \includegraphics[width=0.95\linewidth]{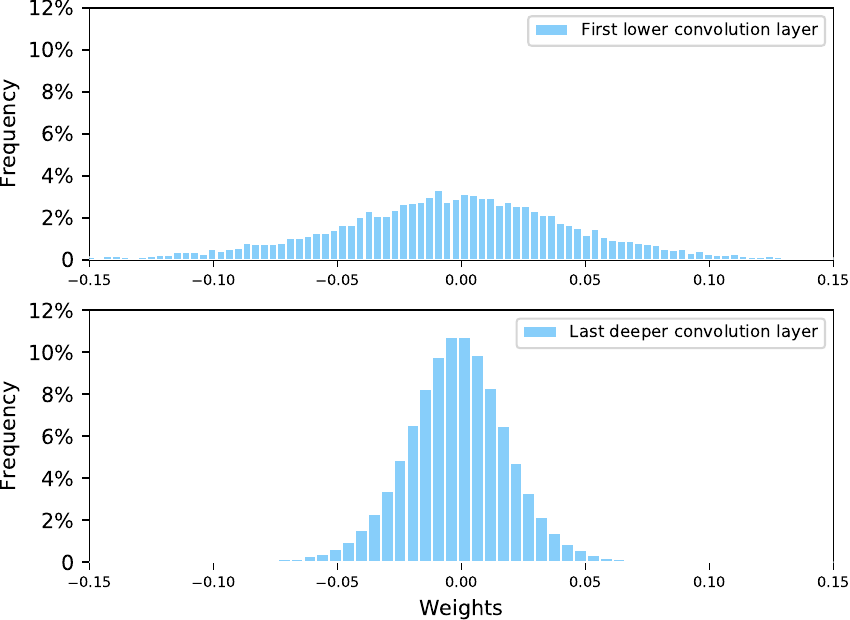}
  \caption{Pre-trained weight distributions for the first lower and last deeper convolutional layers in ResNet34.}
  \label{fig:pre_train}
\end{figure}

As Section \ref{ssec:kmqat} discusses, the shape of real-valued weight distribution differs among various layers in a network. Take ResNet as an example, it is evident from Fig. \ref{fig:pre_train} that distribution in shallow layers displays a broader range and greater variance. In contrast, deeper layers tend to have a more compact and condensed shape. Employing a fixed binary set in static quantization restricts the representational diversity of quantized network. In this section, we present an adaptive quantization approach which is capable of dynamically determining the ideal binary set for each layer. Therefore, it can achieve a better alignment with the distribution of real-valued weights.

Unlike static quantization that utilizes a fixed binary integer set $\left\{-1, +1\right\}$, we introduce two adaptive parameters, $d$ and $\beta$, to dynamically generate different binary sets for each layer, as Figure \ref{fig:binary_quantizer} (right) displays. The binarized weight can be derived using the following equation:
\begin{equation}
  Q = \begin{cases}
    b_1 = \beta - d, & w < \beta\\
    b_2 = \beta + d, & w > \beta
  \end{cases}
  \label{eq13}
\end{equation}
where $\beta$ is defined as the center of binarized weights. $d$ signifies the deviation from this center. In this case, the binary set contains $\left\{b_1, b_2 \right\}$, which varies across different layers.

Next, we will discuss how to determine $d$ and $\beta$ for the $i$-th layer. Starting with weight matrix $W_i$, we initially align the center of binary value $\beta$ with the mean of $W_i$. As a result, $\beta$ can be obtained via:
\begin{equation}
  \beta = \frac{1}{h \times k \times k}\sum_{j=0}^{h-1}\sum_{m=0}^{k-1} \sum_{n=0}^{k-1} W_i^{j,m,n}
  \label{eq14}
\end{equation}
where $h$ and $k$ correspond to channel number and kernel size respectively. $j$, $m$ and $n$ are indices for channel $h$, the first and second dimension of kernel size $k$ respectively.

When estimating $d$, Kullback-Leibler divergence (KLD) is utilized to evaluate the similarity between the distributions of binarized and real-valued weights.
\begin{equation}
  D_{KL}(P_r \parallel P_b) = \int P_r(x)\log\frac{P_r(x)}{P_b(x)} \,dx
  \label{eq15}
\end{equation}
where $P_r(x)$ and $P_b(x)$ represent the probability distributions of real-valued weights and binarized weights respectively.

We assume that binarized weights adhere to a uniform distribution, which means $P_b(b_1)=P_b(b_2)=1/2$. For real-valued weights, it is typically considered that they follow a Gaussian distribution. To minimize the KL distance in Eq. \ref{eq15}, our empirical observations suggest that $d$ should be at the standard deviation of $W_i$, which can be calculated as follows:
\begin{equation}
  d = \frac{\lVert W_i - \beta \rVert_{2}}{\sqrt{h \times k \times k}}
  \label{eq16}
\end{equation}

During quantization-aware training, our adaptive quantization method enables dynamic update of $d$ and $\beta$ along with real-valued weights within each layer.

\section{Experimental Setups}
\subsection{Datasets and Data Augmentation}
In the experiments, we evaluate the proposed methods on Voxceleb1\&2~\cite{voxceleb1, voxceleb2}, which are large-scale and popular benchmark datasets collected for speaker recognition tasks. They comprise interview audio recordings of more than 6000 celebrities, sourced from YouTube website. Specifically, all systems are trained using Voxceleb2 dev set, which contains approximately 2,200 hours of data, featuring a total of 1,092,009 utterances from 5,994 different speakers. During testing, performance is measured on the three official trials: Vox1-O, Vox1-E and Vox1-H. In addition, to enrich the diversity of training data, we employ three different data augmentation techniques, as outlined below:

\begin{itemize}
    \item Speed Perturbation~\cite{speed_perturb}: Sox is used to adjust speech speed by either 0.9 or 1.1 times, increasing speaker number to $17,982$ and training utterances to $3,276,027$.
    \item Online Speech Augmentation~\cite{online_data_aug}: We randomly add the background noise and reverberation from MUSAN~\cite{musan} and RIR~\cite{rir} datasets to training utterances in an online manner. The probability is set to 0.6.
    \item SpecAugment~\cite{specaug}: In addition, random masking is applied to  acoustic features in both frequency and time dimensions.
\end{itemize}

\begin{table*}[ht]
  \caption{EER and MinDCF results of full-precision and quantized ResNets on the Voxceleb1 dataset. ``Uniform'' is fixed uniform precision quantization. ``KMQAT'' represents our proposed adaptive uniform precision quantization. ``MSFT'' denotes multi-stage fine-tuning strategy in mixed precision quantization where ``$\left \{2, 3, 4 \right \}$'' and ``$\left \{1, 2, 3, 4 \right \}$'' refer to the candidate bit set. ``STATIC'' and ``ADAPTIVE'' are the static and adaptive binary quantizers respectively.}
  \label{table:3}
  \centering
  \setlength{\doublerulesep}{4.5pt}
  \begin{adjustbox}{width=.98\textwidth,center}
  \begin{tabular}{|l|c|c|c|c|@{\extracolsep{4pt}}cc|@{\extracolsep{4pt}}cc|@{\extracolsep{4pt}}cc|@{\extracolsep{4pt}}}
    \hline
    \multirow{2}{*}{\makecell{\textbf{System}}}  & \multirow{2}{*}{\makecell{\textbf{Quantization} \\ \textbf{Type}}} & \multirow{2}{*}{\makecell{\textbf{Bit Width} \\ \textbf{(bit)}}} & \multirow{2}{*}{\makecell{\textbf{Model} \\ \textbf{Size}}} & \multirow{2}{*}{\makecell{\textbf{Compression} \\ \textbf{Ratio}}} & \multicolumn{2}{c|}{\textbf{Voxceleb-O}}     & \multicolumn{2}{c|}{\textbf{Voxceleb-E}} & \multicolumn{2}{c|}{\textbf{Voxceleb-H}} \\
    & & & & & \textbf{EER(\%)} & \textbf{MinDCF}  & \textbf{EER(\%)} & \textbf{MinDCF} & \textbf{EER(\%)} & \textbf{MinDCF}    \\
    \hline
    \hline
    ResNet34  & -- & 32 & 26.66MB & -- & 0.888 &  0.0980 & 1.008 & 0.1206 & 1.850 & 0.1837 \\
    \hline
    \multirow{4}{*}{\makecell{+Uniform}}  & \multirow{4}{*}{\makecell{Uniform}} & 4 & 3.45MB & 7.72x & 0.909 &  0.0951 & 1.049 & 0.1251 & 1.923 & 0.1818 \\	
    &  & 3 & 2.63MB & 10.11x & 1.112&  0.1273 & 1.239 & 0.1452 & 2.226 & 0.2077 \\
    &  & 2 & 1.80MB & 14.81x & 1.622& 0.1832 & 1.601 & 0.1871 & 2.845 & 0.2583 \\
    &  & 1 & 0.97MB & 27.48x & 2.143 &  0.2212 & 2.033 & 0.2249 & 3.502 & 0.3016 \\
    \hline
    \multirow{4}{*}{\makecell{+KMQAT}}  & \multirow{4}{*}{\makecell{Uniform}} & 4 & 3.45MB & 7.72x & 0.930 &  0.1068 & 1.033 & 0.1213 & 1.899 & 0.1815 \\	
    &  & 3 & 2.63MB & 10.11x & 0.979&  0.1164 & 1.132 & 0.1282 & 2.028 & 0.1912 \\
    &  & 2 & 1.80MB & 14.81x & 1.319& 0.1375 & 1.306 & 0.1583 & 2.386 & 0.2251 \\
    &  & 1 & 0.97MB & 27.48x & 2.133 &  0.2218 & 2.031 & 0.2242 & 3.498 & 0.3011 \\
    \hline
    \multirow{5}{*}{\makecell{++MSFT}} & Mixed {$\left \{2,3,4\right \}$} & ~3 & 2.57MB & 10.37x & 0.930 & 0.1215 &	1.076  & 0.1245 &	1.972 & 0.1861	\\
    \cline{2-11}
    & \multirow{4}{*}{\makecell{Mixed  \ {$\left \{1,2,3,4\right \}$}}}& ~2 & 1.78MB & 14.94x & 1.148 & 0.1244 &	1.287  & 0.1480 &	2.265 & 0.2093	\\
    & & ~1.7 & 1.58MB & 16.87x &1.297  & 0.1441 & 1.394 & 0.1662 &  2.439	 &  0.2280 \\
    & & ~1.3 & 1.28MB & 20.83x & 1.536 & 0.1758 & 1.574  & 0.1752 & 2.751 &0.2446 \\
    & & ~1.1 & 1.07MB & 24.92x & 1.787 & 0.1973 & 1.839  & 0.2012 & 3.157 & 0.2818 \\
    \hline
    +STATIC & \multirow{2}{*}{\makecell{Binary}} & 1 & 0.97MB & 27.48x &1.902 & 0.2124 & 1.996 & 0.2152 & 3.404 & 0.2981 \\
    \cline{1-1}
    \cline{3-11}
    +ADAPTIVE & & 1 & 0.97MB & 27.48x &  1.725 & 0.2003 & 1.814 & 0.1970 & 3.146 & 0.2782 \\
    \hline
    \hline
    			
    ResNet101  & -- & 32 &   63.86MB &  --  & 0.670 &  0.0681&  0.777& 0.0843 & 1.440 & 0.1327 \\
    \hline
    \multirow{4}{*}{\makecell{+Uniform}}  & \multirow{4}{*}{\makecell{Uniform}} & 4 & 8.42MB & 7.58x & 0.675 &  0.0781 & 0.913 & 0.0974 & 1.569 & 0.1432 \\	
    &  & 3 & 6.44MB & 9.91x & 0.798&  0.0874 & 0.943 & 0.1065 & 1.751 & 0.1605 \\
    &  & 2 & 4.47MB & 14.31x & 1.005& 0.1178 & 1.212 & 0.1334 & 2.012 & 0.1897 \\
    &  & 1 & 2.48MB & 25.75x & 1.279 &  0.1382 & 1.353 & 0.1495 & 2.367 & 0.2150 \\
    \hline
    \multirow{4}{*}{\makecell{+KMQAT}}  & \multirow{4}{*}{\makecell{Uniform}} & 4 & 8.42MB & 7.58x & 0.670 & 0.0763 & 0.826 & 0.0927 & 1.511 & 0.1403 \\	
    &  & 3 & 6.44MB & 9.91x & 0.713&  0.0795 & 0.866 & 0.0969 & 1.579 & 0.1473 \\
    &  & 2 & 4.47MB & 14.31x & 0.824& 0.0909 & 0.995 & 0.1131 & 1.809 & 0.1666 \\
    &  & 1 & 2.48MB & 25.75x & 1.271 &  0.1368 & 1.350 & 0.1490 & 2.363 & 0.2147 \\
    \hline
    \multirow{5}{*}{\makecell{++MSFT}} & Mixed {$\left \{2,3,4\right \}$} & ~3 & 6.39MB & 9.99x & 0.670 & 0.0855&	0.865  & 0.0962 &	1.578 & 	0.1488\\
    \cline{2-11}
    & \multirow{4}{*}{\makecell{Mixed  \ {$\left \{1,2,3,4\right \}$}}}& ~2 & 4.39MB & 14.55x & 0.830 & 0.0971 &	0.944  & 0.1090 & 1.678 & 0.1576	\\
    & & ~1.7 & 3.69MB & 17.31x & 0.877&0.0971 &1.026  & 0.1147 &  1.785	 & 0.1625  \\
    & & ~1.3 & 3.09MB & 20.67x & 1.058 & 0.1027 & 1.098  & 0.1252 & 1.962 & 0.1821\\
    & & ~1.1 & 2.69MB & 23.73x & 1.138 & 0.1145 & 1.211 & 0.1340 & 2.127 &0.1959 \\
    \hline
    +STATIC & \multirow{2}{*}{\makecell{Binary}} & 1 & 2.48MB & 25.75x &1.203 & 0.1291 & 1.275 & 0.1421 & 2.237 & 0.2085 \\
    \cline{1-1}
    \cline{3-11}
    +ADAPTIVE & & 1 & 2.48MB & 25.75x &  1.114 & 0.1160 & 1.174 & 0.1326 & 2.093 & 0.1931 \\
    \hline
  \end{tabular}
  \end{adjustbox}
\end{table*}

\subsection{System Description}
In this paper, ResNet-based SV systems are implemented as the baselines, which are the most commonly adopted networks and provide state-of-the-art performance. Specifically, two different types of architectures are employed, as detailed below:

\begin{itemize}
    \item ResNet~\cite{rvector}: It serves as a strong and powerful model in the SV field. We build ResNet34 and ResNet101 as full-precision systems in the following experiments.
    \item DF-ResNet~\cite{df-resnet}: This is an enhanced version of ResNet. Similarly, DF-ResNet110 and DF-ResNet179 are included as the baseline models.
\end{itemize}

\subsection{Training Strategies}
\label{sssec:training}
The entire training process contains two stages. In the first stage, we perform full-precision pre-training on ResNets and DF-ResNets systems. Subsequently, we apply the proposed adaptive uniform precision, mixed precision and binary quantization methods to the pre-trained networks respectively, resulting in the corresponding quantized models.

\textbf{stage 1: }We firstly pre-train ResNets and DF-ResNets with full precision. For each training utterance, we randomly chunk a 200-frame segment during the pre-training process. Then, 80-dimensional filter bank is extracted as acoustic features using 25ms window length and 10ms hop size. Plus, loss function is AAM-softmax~\cite{sv-loss4} with a 0.2 margin and a scale of 32. ResNets adopt SGD optimizer with a momentum of 0.9 and a weight decay of 1e-4, while DF-ResNets are optimized using AdamW~\cite{adamw} with a weight decay of 0.05. The dimension of speaker embedding is set to 256.

\textbf{stage 2: }Then, the pre-trained full-precision systems are quantized and fine-tuned through our proposed approaches. During quantization-aware training, online speech augmentation and SpecAugment are omitted. Specifically, we perform 40 epochs of retraining for ResNets and 80 epochs for DF-ResNets. The remaining settings are consistent with stage 1.

\subsection{Evaluation Metrics}
For testing, similarity score between speaker embeddings is measured using cosine distance. In addition, we perform adaptive score normalization (AS-Norm)~\cite{asnorm} post-processing with an imposter cohort size of 600. Performance is reported on equal error rate (EER) and minimum detection cost function (MinDCF) with the settings of $P_{target} = 0.01$ and $C_{FA}=C_{Miss}=1$.

\section{Results and Analysis}
This part starts with the evaluation of our adaptive uniform precision quantization in Section \ref{sssec:uniform_results}. Subsequently, we present an in-depth discussion of mixed precision quantization results in Section \ref{sssec:mixed_results}. Section \ref{sssec:binary_results} then enumerates the outcomes of two specially designed 1-bit quantization schemes. The detailed analysis of weight distribution in quantized models is covered in Section \ref{sssec:weight_analysis}. Finally, Section \ref{sssec:comparison} provides a thorough comparison of our newly developed model families with various existing lightweight SV systems.

\begin{table*}[ht]
  \caption{EER and MinDCF results of full-precision and quantized DF-ResNets on the Voxceleb1 dataset. The experimental settings are the same as ResNets in Table \ref{table:3}.}
  \label{table:4}
  \centering
  \setlength{\doublerulesep}{4.5pt}
  \begin{adjustbox}{width=.98\textwidth,center}
  \begin{tabular}{|l|c|c|c|c|@{\extracolsep{4pt}}cc|@{\extracolsep{4pt}}cc|@{\extracolsep{4pt}}cc|@{\extracolsep{4pt}}}
    \hline
    \multirow{2}{*}{\makecell{\textbf{System}}}  & \multirow{2}{*}{\makecell{\textbf{Quantization} \\ \textbf{Type}}} & \multirow{2}{*}{\makecell{\textbf{Bit Width} \\ \textbf{(bit)}}} & \multirow{2}{*}{\makecell{\textbf{Model} \\ \textbf{Size}}} & \multirow{2}{*}{\makecell{\textbf{Compression} \\ \textbf{Ratio}}} & \multicolumn{2}{c|}{\textbf{Voxceleb-O}}     & \multicolumn{2}{c|}{\textbf{Voxceleb-E}} & \multicolumn{2}{c|}{\textbf{Voxceleb-H}} \\
    & & & & & \textbf{EER(\%)} & \textbf{MinDCF}  & \textbf{EER(\%)} & \textbf{MinDCF} & \textbf{EER(\%)} & \textbf{MinDCF}    \\
    \hline
    \hline
    DF-ResNet110  & -- & 32 & 29.12MB & -- & 0.713 & 0.0542  & 0.863 & 0.0968 & 1.583 & 0.1511 \\
    \hline
    \multirow{4}{*}{\makecell{+Uniform}} & \multirow{4}{*}{\makecell{Uniform}} & 4 & 4.30MB & 6.77x & 0.685 &  0.0693& 0.913 & 0.1043 & 1.667& 0.1578 \\	
    &  & 3 & 3.40MB & 8.56x & 0.991& 0.0910 & 1.115 & 0.1260 & 1.997 & 0.1905 \\
    &  & 2 & 2.51MB & 11.60x & 1.157& 0.1406 & 1.339 & 0.1836 & 2.358 & 0.2611 \\
    &  & 1 & 1.62MB & 17.98x & 2.241 &  0.2119 & 2.285 & 0.2322 & 3.702 & 0.3174 \\
    \hline
    \multirow{4}{*}{\makecell{+KMQAT}} & \multirow{4}{*}{\makecell{Uniform}} & 4 & 4.30MB & 6.77x & 0.681 &  0.0689& 0.898 & 0.1025 & 1.639& 0.1571 \\	
    &  & 3 & 3.40MB & 8.56x & 0.894& 0.0835 & 0.996 & 0.1125 & 1.818 & 0.1717 \\
    &  & 2 & 2.51MB & 11.60x & 0.957& 0.1099 & 1.126 & 0.1583 & 2.016 & 0.2251 \\
    &  & 1 & 1.62MB & 17.98x & 2.238 &  0.2114 & 2.281 & 0.2319 & 3.699 & 0.3166 \\
    \hline
    \multirow{5}{*}{\makecell{++MSFT}} & Mixed {$\left \{2,3,4\right \}$} & ~3 & 3.36MB&8.65x & 0.766 & 0.0779 &	0.923  & 0.1065 &	1.717& 0.1580	\\
    \cline{2-11}
    & \multirow{4}{*}{\makecell{Mixed  \ {$\left \{1,2,3,4\right \}$}}}& ~2 & 2.50MB&11.69x &0.896 & 0.0762 &	1.061  & 0.1204 &	1.885 & 0.1887	\\
    & & ~1.7 & 2.25MB&12.96x &1.147  & 0.1239 & 1.241 & 0.1376 &  2.137	 &  0.1985 \\
    & & ~1.3 & 1.92MB&15.12x & 1.334 & 0.1452 & 1.514  & 0.1631 & 2.626 &0.2422 \\
    & & ~1.1 & 1.72MB&16.94x &1.776 & 0.1939 & 1.838  & 0.1972 & 3.177 & 0.2813\\
    \hline
    +STATIC & \multirow{2}{*}{\makecell{Binary}} & 1 & 1.62MB & 17.98x & 1.873 &  0.2038 & 2.024 & 0.2107 & 3.341 & 0.2995 \\
    \cline{1-1}
    \cline{3-11}
    +ADAPTIVE &  & 1 & 1.62MB & 17.98x & 1.702 &  0.1971 & 1.805 & 0.1943 & 3.164 & 0.2778 \\
    \hline
    \hline
    DF-ResNet179  & -- & 32 &   39.97MB & -- & 0.622 & 0.0678& 0.799& 0.0875 & 1.452 & 0.1411\\
    \hline
    \multirow{4}{*}{\makecell{+Uniform}}  & \multirow{4}{*}{\makecell{Uniform}} & 4 & 5.89MB & 6.69x & 0.664 & 0.0657 &0.863 &0.0952 & 1.551 & 0.1506 \\	
    &  & 3 & 4.76MB & 8.39x & 0.7303&  0.0819 & 1.007 & 0.1073 & 1.824 & 0.1682 \\
    &  & 2 & 3.55MB & 11.26x & 1.096& 0.1215 & 1.229 & 0.1432 & 2.157 & 0.2119 \\
    &  & 1 & 2.33MB & 17.11x & 1.870 &  0.2141 &1.953 & 0.2120 & 3.228 & 0.2859 \\
    \hline
    \multirow{4}{*}{\makecell{+KMQAT}}  & \multirow{4}{*}{\makecell{Uniform}} & 4 & 5.89MB & 6.69x & 0.670 & 0.0668 &0.858 &0.0933 & 1.548 & 0.1494 \\	
    &  & 3 & 4.76MB & 8.39x & 0.670&  0.0738 & 0.892 & 0.0994 & 1.615 & 0.1544 \\
    &  & 2 & 3.55MB & 11.26x & 0.899& 0.0948 & 1.042 & 0.1194 & 1.842 & 0.1766 \\
    &  & 1 & 2.33MB & 17.11x & 1.867 &  0.2135 &1.949 & 0.2116 & 3.223 & 0.2852 \\
    \hline
    \multirow{5}{*}{\makecell{++MSFT}} & Mixed {$\left \{2,3,4\right \}$} & ~3 & 4.70MB&8.51x & 0.611 & 0.0763&	0.891  & 0.1051 &	1.616 & 0.1516	\\
    \cline{2-11}
    & \multirow{4}{*}{\makecell{Mixed  \ {$\left \{1,2,3,4\right \}$}}}& ~2 & 3.55MB&11.26x & 0.803 & 0.0945 &	0.969  & 0.1084 & 1.723 & 0.1647	\\
    & & ~1.7 & 3.17MB&12.64x & 0.921&0.1101 &1.103  & 0.1238 &  1.938	 & 0.1890  \\
    & & ~1.3 & 2.70MB&14.83x & 1.228 & 0.1493 & 1.391  & 0.1532 & 2.371 & 0.2247\\
    & & ~1.1 & 2.46MB&16.21x & 1.329 & 0.1416 & 1.531 & 0.1747 & 2.695& 0.2505 \\
    \hline
    +STATIC & \multirow{2}{*}{\makecell{Binary}} & 1 & 2.33MB & 17.11x & 1.516 &  0.1643 & 1.715 & 0.1862 & 2.931 & 0.2654 \\
    \cline{1-1}
    \cline{3-11}
    +ADAPTIVE &  & 1 & 2.33MB & 17.11x & 1.275 &  0.1472 & 1.494 & 0.1713 & 2.636 & 0.2471 \\
    \hline
  \end{tabular}
  \end{adjustbox}
\end{table*}

\subsection{Evaluation on Adaptive Uniform Precision Quantization}
\label{sssec:uniform_results}
We firstly examine the performance of our adaptive uniform precision quantization approach, namely KMQAT, on both ResNets and DF-ResNets.

\begin{figure*}[ht]
\begin{minipage}{0.5\textwidth}
\centering
  \begin{subfigure}[t]{.98\textwidth}
    \centering
    \includegraphics[width=1.0\textwidth]{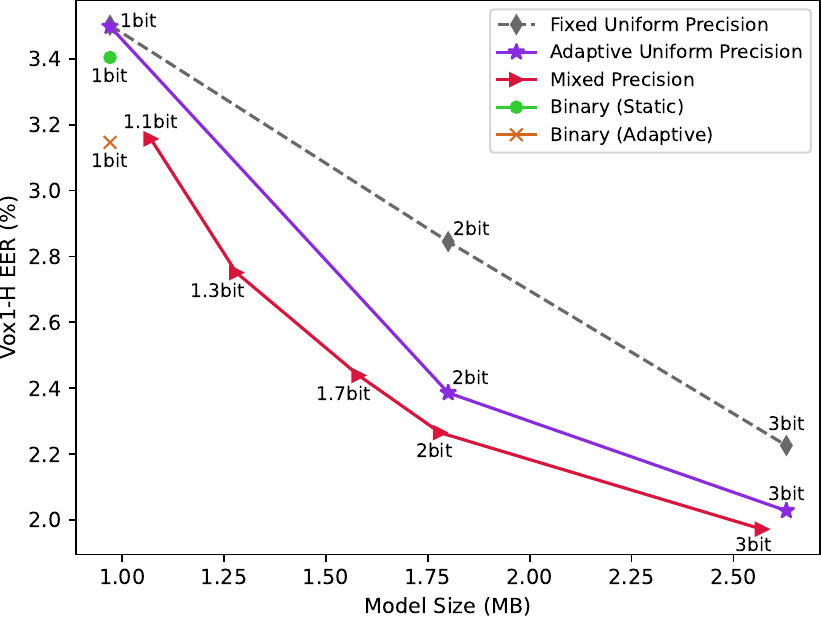}
    \caption{ResNet34}
  \end{subfigure}\vspace{1em}
  \begin{subfigure}[t]{.98\textwidth}
    \centering
    \includegraphics[width=1.0\textwidth]{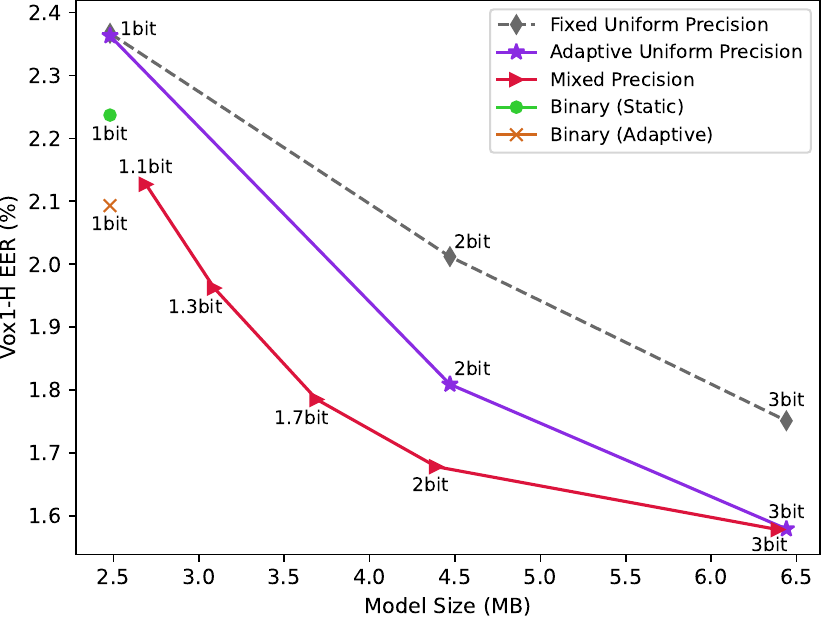}
    \caption{ResNet101}
  \end{subfigure}
\end{minipage}%
\begin{minipage}{0.5\textwidth}
\centering
  \begin{subfigure}[t]{.98\textwidth}
    \centering
    \includegraphics[width=1.0\textwidth]{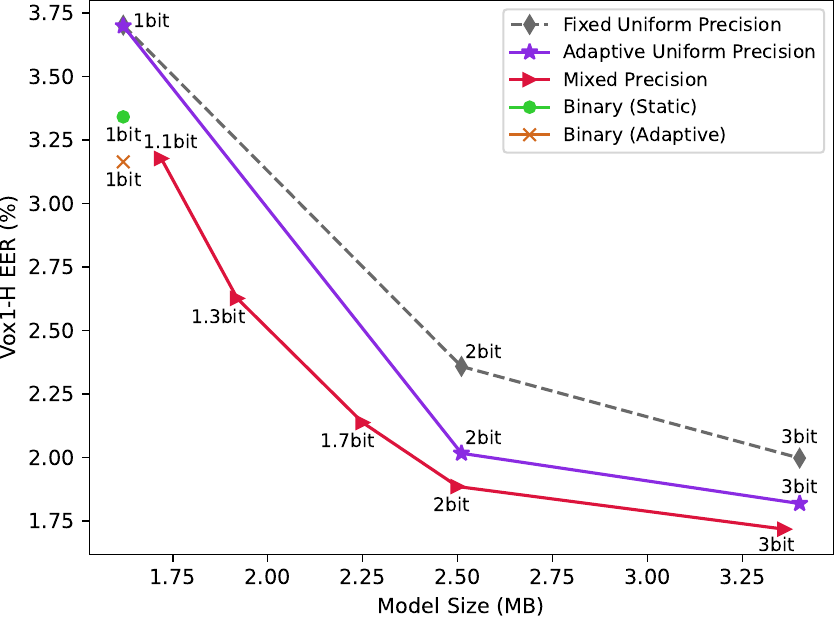}
    \caption{DF-ResNet110}
  \end{subfigure}\vspace{1em}
  \begin{subfigure}[t]{.98\textwidth}
    \centering
    \includegraphics[width=1.0\textwidth]{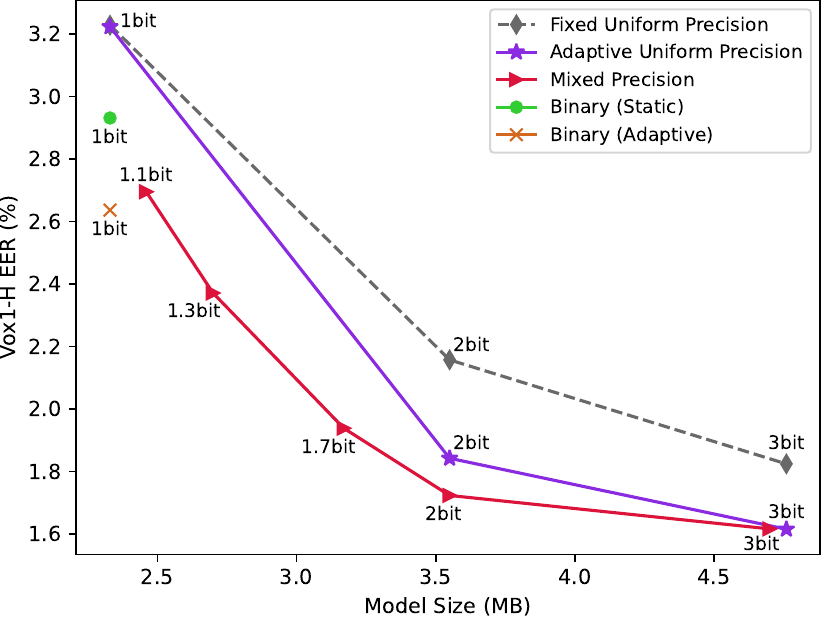}
    \caption{DF-ResNet179}
  \end{subfigure}
\end{minipage}

\caption{The comparison of fixed uniform precision, adaptive uniform precision, mixed precision and binary quantization results in terms of performance and model size on ResNets and DF-ResNets.}
\label{fig:comparison}
\end{figure*}

The results presented in Table \ref{table:3} and \ref{table:4} demonstrate that KMQAT effectively realizes lossless 4-bit uniform precision quantization on both ResNets and DF-ResNets, offering an impressive compression ratio of approximately $\sim$8 and $\sim$7 respectively. For example, when applying KMQAT to ResNet34 and DF-ResNet110, the resulting 4-bit quantized models attain nearly identical performance as opposed to their full-precision versions. This confirms the effectiveness of using k-means clustering for the generation of quantization centroids. Similarly, the performance loss caused by 4-bit compression in larger and deeper networks, such as ResNet101 and DF-ResNet179, is negligible. This highlights the strong generality of our proposed KMQAT method, suggesting its broad applicability across different network architectures. Compared to fixed uniform method, our adaptive one achieves much better performance across various bit precisions, as Fig. \ref{fig:comparison} shows.

In addition, when bit width is reduced, while compression ratio becomes higher, the performance of quantized models will severely degrade. Specifically, in the cases of 3-bit and 2-bit, ResNets and DF-ResNets suffer an average performance reduction of 10.1\% and 21.4\% respectively. For 1-bit scenario, the compression ratio reaches its peak, approximately 27x for ResNets and 18x for DF-ResNets. However, this comes at the cost of the worst EER, which is twice as poor as that of full-precision systems. This phenomenon reveals that the representation capacity of uniform precision quantization at low bits still remains limited.





\subsection{Evaluation on Mixed Precision Quantization}
\label{sssec:mixed_results}
As mentioned above, employing low-bit compression within uniform precision quantization can lead to considerable performance degradation. This section further investigates the effect of the introduced mixed precision quantization method. As outlined in Section \ref{sec:mixed_precision}, the initial step of our mixed precision quantization method involves identifying a set of possible bits for a target model size. In the experiments, we set candidate set $\mathcal{C}$ as $\{2, 3, 4\}$ to achieve a mixed precision quantization equivalent to 3-bit and $\{1, 2, 3, 4\}$ for cases below 3-bit.

From Table \ref{table:3} and \ref{table:4}, it can be clearly observed that the proposed mixed precision approach significantly outperforms uniform precision quantization, particularly in low-bit scenarios, while maintaining a similar model size and compression ratio. Specifically, bit combinations discovered for ResNet34 and DF-ResNet110 achieve the same level of 3-bit quantization as uniform precision with a slightly smaller model size. When integrated with MSFT, these mixed precision quantized models exhibit superior performance. Similarly, applying 2-bit mixed precision quantization to ResNets and DF-ResNets obtains average performance improvements of 6.2\% and 6.5\%, respectively. For cases below 2-bit, we explore three additional configurations: 1.7-bit, 1.3-bit and 1.1-bit. In these contexts, mixed precision quantization demonstrates a larger advantage compared to KMQAT. For instance, 1.1-bit equivalent quantization surpasses the performance of 1-bit KMQAT by an average of 14\% with only a slight increase in model size. In addition, Fig. \ref{fig:comparison} presents a schematic comparison between uniform and mixed precision quantization, illustrating that the latter achieves a better trade-off on performance and model size for regime below 3-bit. The above findings suggest that the searched bit combination based on Hessian sensitivity analysis is more effective and reasonable than uniform one, thereby boosting the performance of low-bit quantized models. Moreover, MSFT proves to be a powerful strategy for incrementally quanitzing and fine-tuning network in the context of mixed precision quantization.

Another benefit of mixed precision quantization is its ability to create suitable bit combination for any desirable model size, whereas uniform precision quantization is limited to producing only integer bit quantized models. As Fig. \ref{fig:comparison} shows, we can attain models quantized equivalently at either 1.7 bits or 1.3 bits through mixed precision quantization, which is not feasible with uniform quantization.

\begin{figure}[!t]
  \centering
  \includegraphics[width=0.95\linewidth]{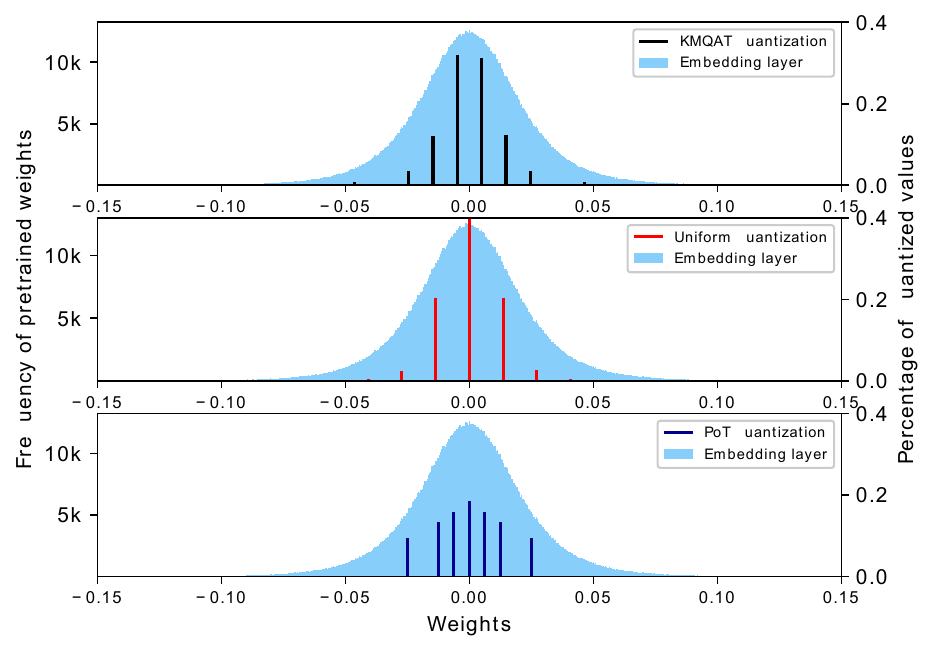}
  \caption{Pre-trained weight distribution and 3-bit quantization levels for the embedding layer in ResNet34.}
  \label{fig:kmqat}
\end{figure}

\subsection{Evaluation on Binary Quantization}
\label{sssec:binary_results}
In this section, we delve into the results of two specially designed 1-bit quantization schemes: the static and adaptive binary quantizers.

As mentioned earlier, although 1-bit quantization offers the highest compression, it also causes the performance of quanitzed models to drop to their lowest. It is evident from Table \ref{table:3} and \ref{table:4} that both static and adaptive quantization schemes can effectively recover the performance of binarized models. Specifically, compared to 1-bit KMQAT, our static 1-bit quantizer obtains obvious performance gains, averaging 7.7\% for ResNets and 9.4\% for DF-ResNets. Furthermore, the adaptive quantizer exhibits superior performance over static version, even surpassing the performance of 1.1-bit mixed precision quantiation, as Fig. \ref{fig:comparison} illustrates. The above analyses indicate that two specially designed 1-bit quantization schemes can enhance the representational capacity of binarized models, and the adaptive method is more effective than the static one.

\subsection{Weight Distribution Analysis}
\label{sssec:weight_analysis}
To better verify the efficiency of our proposed quantization approaches, this section provides a visualization and analysis of weight distribution in the quantized ResNet34.

\subsubsection{Adaptive Uniform Precision Quantization}
We initially evaluate the proposed adaptive uniform precision quantization by contrasting its quantization levels with those of earlier methods. From Fig. \ref{fig:kmqat}, we can clearly see that KMQAT more closely aligns with the distribution of pre-trained weights compared to Uniform and PoT methods. For example, Uniform quantization assigns fewer levels to the peak area, whereas PoT ignores weights in the edge area. Such an imbalanced distribution of quantization levels could degrade performance. By comparison, KMQAT accounts for all positions in weight distribution, yielding more reasonable quantization levels in both peak and edge areas.

\begin{figure}[!t]
  \centering
  \includegraphics[width=0.95\linewidth]{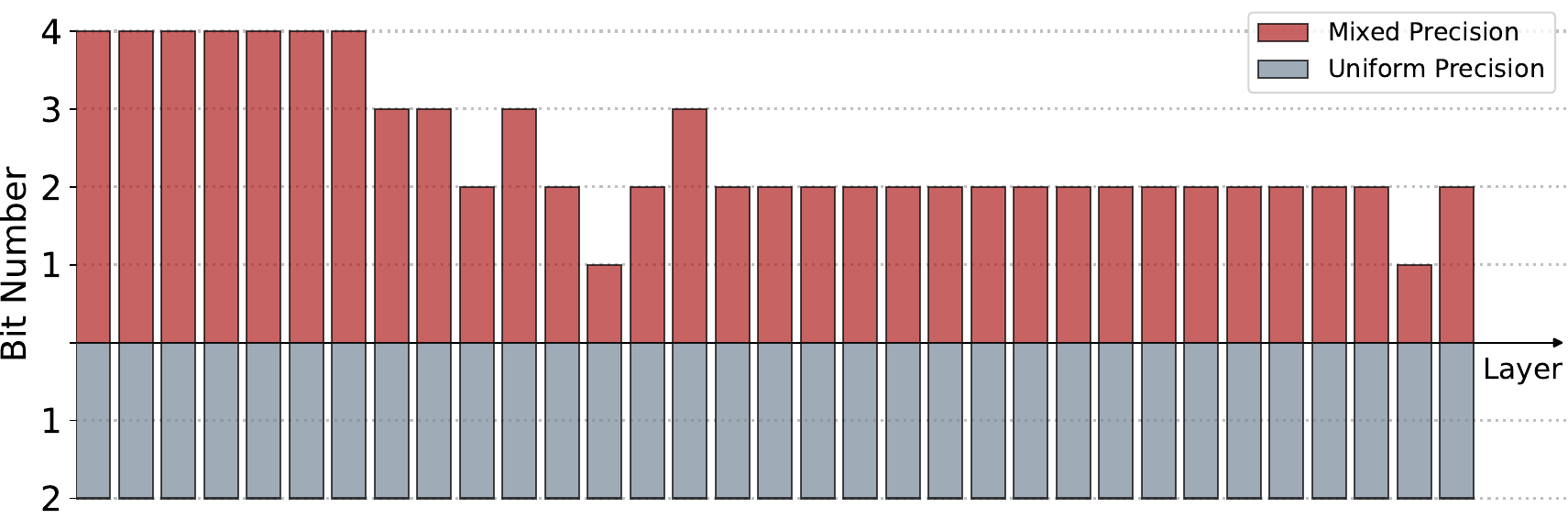}
  \caption{The comparison of bit allocation for each network layer between 2-bit uniform and mixed precision quantization in ResNet34.}
  \label{fig:mixed}
\end{figure}


\begin{figure}[!t]
  \centering
  \includegraphics[width=0.95\linewidth]{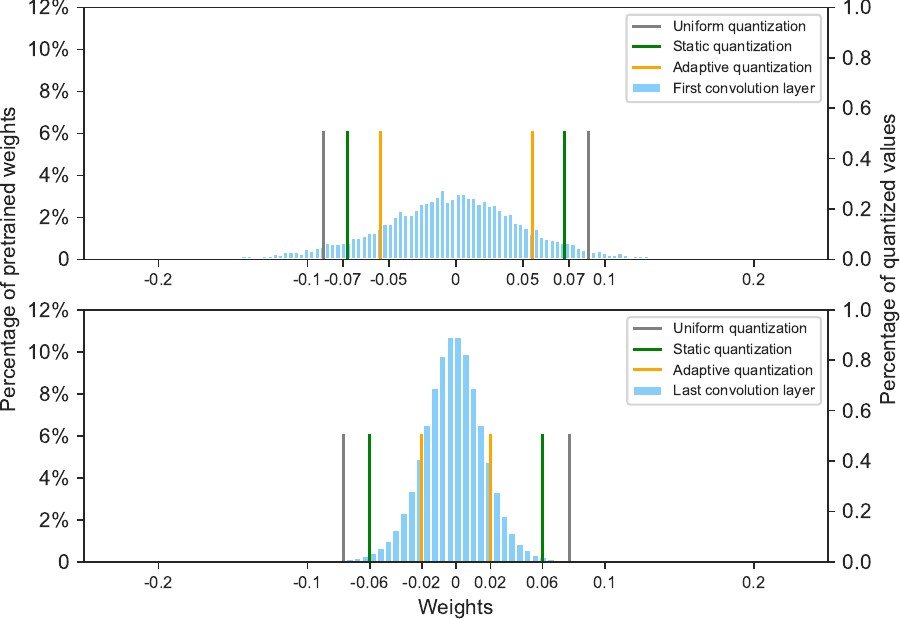}
  \caption{Pre-trained and binarized weight distributions for the first lower and last deeper convolutional layers in ResNet34.}
  \label{fig:binary_weight}
\end{figure}

\subsubsection{Mixed Precision Quantization}
In this part, we illustrate the bit combination found for each layer when applying 2-bit mixed precision quantization. Taking ResNet34 as an example, the candidate bit set is $\{1, 2, 3, 4\}$. As Fig. \ref{fig:mixed} shows, uniform precision quantization uses a fixed 2-bit width for all layers. On the contrary, our proposed mixed precision quantization enables the assignment of different bit widths to each layer. Specifically, shallower convolutional layers demand higher precision, typically 4 bits, as they process raw and low-level data where low precision could lead to substantial information loss. Conversely, deeper layers, dealing with dense and abstract features, can maintain effective performance even with lower precision, such as 1 or 2 bits, in quantized models. This confirms that mixed precision strategy can yield a more effective and reasonable bit allocation than the uniform approach.

\subsubsection{Binary Quantization}
Finally, we analyze the binarized weight histograms of uniform, static and adaptive quantization schemes. The distributions of pre-trained weights, as shown in Fig. \ref{fig:binary_weight}, display a notable disparity between the first and last convolutional layers. It is evident that uniform quantization gives the worst binarized results, failing to align with the real-valued weight distribution. Using a fixed integer set, static 1-bit quantization offers a closer match to the pre-trained weight distribution. However, it produces nearly identical binarized outcomes in the first and last layers ($0.07$ vs. $0.06$). For example, most weights in the final layer fall within the range $[-0.04, 0.04]$, yet static quantization yields two binarized weights at $\pm0.06$. In contrast, adaptive quantization can determine binary values for different layers according to their distributions, achieving the best performance. In the broader first layer, weights are mapped to $\pm0.05$. Meanwhile, for the last layer with a more compact shape, it generates binarized weights of $-0.02$ and $0.02$. This illustrates the superiority of adaptive scheme over both the uniform and static method.

\subsection{Comparison with Other Lightweight Systems}
\label{sssec:comparison}
In this section, we present a thorough comparison between our proposed models and a variety of existing lightweight SV systems. An extensive review of different lightweight approaches are included, such as knowledge distillation, network quantization, and efficient architectural designs.

Table~\ref{table:5} clearly demonstrates that our resulting models surpass previously published lightweight SV systems in performance across all model size ranges. To ensure a fair comparison, we firstly re-implement existing quantization techniques such as PoT~\cite{pot}, APoT~\cite{apot}, ADMM~\cite{admm} and TWN~\cite{twn}. Our developed methods, including adaptive uniform precision and mixed precision quantization, exhibit superior performance, particularly at lower bits, while maintaining similar model sizes. For example, DF-ResNet179 with 2-bit mixed quantization achieves relative EER improvements of up to 50\% over PoT, APoT and ADMM at 4 or 6 bits. For 2 and 3-bit quantization, our KMQAT and mixed quantized models outperform APoT and TWN by up to 21\% with smaller model sizes.

In addition, it is noted that our quantized models achieve much better performance than those based on knowledge distillation and efficient architecture designs. Specifically, ResNet34-KMQAT surpasses Thin and Fast-ResNet34 by a large margin. Moreover, it obtains 28\% relative performance gains compared to ResNet34-SKDFE while being 6.9x smaller in model size. This further validates the effectiveness of our methods.

In the realm of extreme quantization, KMQAT and two advanced 1-bit systems attain a new state-of-the-art performance. Our best system ResNet34-Adaptive achieves an average relative improvement by 35\% compared to CS-CTCSConv1D and ECAPA-TDNNLite, while maintaining a nearly identical model size. Though~\cite{binary-lm} introduces a binary version of ResNet34 with a model size of 0.66MB, its performance significantly lags behind that of our binarized models.

\begin{table}[t]
  \footnotesize
  \caption{Performance comparison between our proposed models and various existing lightweight SV systems on Vox1-H. Model size and bit width are presented in detail.}
  \label{table:5}
  \centering
  \begin{adjustbox}{width=.48\textwidth,center}
  \begin{tabular}{lccc@{\extracolsep{3pt}}}
    \toprule
    \textbf{System} & {\makecell{\textbf{Model Size} \\ \textbf{(MB)}} } & {\makecell{\textbf{Bit Width} \\ \textbf{(bit)}} } & {\makecell{\textbf{Vox1-H} \\ \textbf{EER(\%)}} } \\
    \midrule
    \textbf{DF-ResNet179-Mixed} & 3.55 & 2 & \textbf{1.72}  \\
    \textbf{ResNet101-Mixed} & 3.69 & \textbf{1.7} & \textbf{1.78}  \\
    \textbf{ResNet34-KMQAT} & \textbf{3.45} & 4 & \textbf{1.89}  \\
    ResNet34-PoT~\cite{pot}(our impl.)  & \textbf{3.45} & 4  & 2.06  \\
    ResNet34-APoT~\cite{apot}(our impl.)  & \textbf{3.45} & 4  & 1.90  \\
    ResNet34-ADMM~\cite{admm}(our impl.) & 5.13 & 6  & 3.32  \\
    Thin-ResNet34~\cite{thin-resnet2}  & 5.6 & 32  & 4.09  \\
    Fast-ResNet34~\cite{fast-resnet}  & 5.6 & 32  & 4.21  \\
    ResNet34-SKDFE~\cite{skd}  & 23.9 & 32  & 2.76  \\
    \midrule
    \textbf{DF-ResNet110-Mixed} & \textbf{2.50} & \textbf{2} & \textbf{1.88}  \\
    \textbf{ResNet34-Mixed} & \textbf{2.57} & 3 & \textbf{1.97}  \\
    \textbf{DF-ResNet110-KMQAT} & \textbf{2.51} & \textbf{2} & \textbf{2.01}  \\
    \textbf{ResNet34-KMQAT} & \textbf{2.63} & 3 & \textbf{2.02}  \\
    ResNet34-APoT~\cite{apot}(our impl.)  & 2.63 & 3  &  2.09 \\
    \midrule
    \textbf{ResNet34-Mixed} & \textbf{1.78} & 2 & \textbf{2.26}  \\
    \textbf{ResNet34-KMQAT} & 1.80 & 2 & \textbf{2.38}  \\
    \textbf{DF-ResNet110-Mixed} & 1.92 & \textbf{1.3} & \textbf{2.62}  \\
    ResNet34-TWN~\cite{twn}(our impl.) & 1.80 & 2 & 2.76  \\
    ResNet34-APoT~\cite{apot}(our impl.)  & 1.80 & 2  &  2.86 \\
    \midrule
    \textbf{ResNet34-Adaptive} & 0.97 & \textbf{1}  & \textbf{3.14}  \\
    \textbf{ResNet34-Static} & 0.97 & \textbf{1}  & \textbf{3.40}  \\
    \textbf{ResNet34-KMQAT} & 0.97 & \textbf{1}  & \textbf{3.78}  \\
    ResNet34-Binary~\cite{binary-lm}  & \textbf{0.66} & \textbf{1}  & 5.35  \\
    CS-CTCSConv1D~\cite{cs-ctcsconv1d} & 0.96 & 32 & 4.44  \\
    ECAPA-TDNNLite~\cite{ecapa-lite} & 1.2 & 32 & 5.20  \\
    \bottomrule
  \end{tabular}
  \end{adjustbox}

\end{table}

\section{Conclusion}
In this paper, we explore adaptive neural network quantization for lightweight speaker verification. Firstly, we introduce an innovative adaptive uniform precision quantization technique that leverages k-means clustering to dynamically generate quantization centroids specific to each network layer. To improve the performance of low-bit quantized models, a new algorithm that merges mixed precision quantization with a multi-stage fine-tuning (MSFT) strategy is further developed. Finally, we propose two distinct binary quantization schemes tailored for 1-bit scenario: the static and adaptive quantizers. Experimental results demonstrate that both ResNets and DF-ResNets effectively attain 4-bit uniform precision quantization with negligible performance loss. Furthermore, compared to the uniform precision approach, mixed precision quantization not only achieves better performance with a similar model size but also enables the generation of bit combination for any desirable model size. In addition, our newly designed 1-bit quantization schemes significantly enhance the performance of binarized models. The visualization of quantized weight distributions validates the superiority of our proposed quantization methods. Finally, an in-depth comparison with existing lightweight SV systems indicates that our resulting models substantially outperform earlier systems across various model size ranges.

\bibliographystyle{IEEEtran}
\bibliography{mybib}

\begin{thebibliography}{10}
\providecommand{\url}[1]{#1}
\csname url@samestyle\endcsname
\providecommand{\newblock}{\relax}
\providecommand{\bibinfo}[2]{#2}
\providecommand{\BIBentrySTDinterwordspacing}{\spaceskip=0pt\relax}
\providecommand{\BIBentryALTinterwordstretchfactor}{4}
\providecommand{\BIBentryALTinterwordspacing}{\spaceskip=\fontdimen2\font plus
\BIBentryALTinterwordstretchfactor\fontdimen3\font minus \fontdimen4\font\relax}
\providecommand{\BIBforeignlanguage}[2]{{%
\expandafter\ifx\csname l@#1\endcsname\relax
\typeout{** WARNING: IEEEtran.bst: No hyphenation pattern has been}%
\typeout{** loaded for the language `#1'. Using the pattern for}%
\typeout{** the default language instead.}%
\else
\language=\csname l@#1\endcsname
\fi
#2}}
\providecommand{\BIBdecl}{\relax}
\BIBdecl

\bibitem{ivector}
N.~Dehak, P.~J. Kenny, R.~Dehak, P.~Dumouchel, and P.~Ouellet, ``Front-end factor analysis for speaker verification,'' \emph{IEEE Transactions on Audio, Speech, and Language Processing}, vol.~19, no.~4, pp. 788--798, 2011.

\bibitem{plda}
S.~Ioffe, ``Probabilistic linear discriminant analysis,'' in \emph{European Conference on Computer Vision (ECCV)}, 2006, pp. 531--542.

\bibitem{tandem}
Y.~Liu, Y.~Qian, N.~Chen, T.~Fu, Y.~Zhang, and K.~Yu, ``Deep feature for text-dependent speaker verification,'' \emph{Speech Communication}, vol.~73, pp. 1--13, 2015.

\bibitem{tdnn-sv}
D.~Snyder, D.~Garcia-Romero, D.~Povey, and S.~Khudanpur, ``Deep neural network embeddings for text-independent speaker verification,'' in \emph{Proc. Interspeech}, 2017, pp. 999--1003.

\bibitem{xvector}
D.~Snyder, D.~Garcia-Romero, G.~Sell, D.~Povey, and S.~Khudanpur, ``X-vectors: Robust dnn embeddings for speaker recognition,'' in \emph{IEEE International Conference on Acoustics, Speech and Signal Processing (ICASSP)}, 2018, pp. 5329--5333.

\bibitem{ext-xvector}
D.~Snyder, D.~Garcia-Romero, G.~Sell, A.~McCree, D.~Povey, and S.~Khudanpur, ``Speaker recognition for multi-speaker conversations using x-vectors,'' in \emph{IEEE International Conference on Acoustics, Speech and Signal Processing (ICASSP)}, 2019, pp. 5796--5800.

\bibitem{rvector}
H.~Zeinali, S.~Wang, A.~Silnova, P.~Mat{\v{e}}jka, and O.~Plchot, ``But system description to voxceleb speaker recognition challenge 2019,'' \emph{arXiv preprint arXiv:1910.12592}, 2019.

\bibitem{d-tdnn}
Y.~Yu and W.~Li, ``Densely connected time delay neural network for speaker verification,'' in \emph{Proc. Interspeech}, 2020, pp. 921--925.

\bibitem{ecapa}
B.~Desplanques, J.~Thienpondt, and K.~Demuynck, ``Ecapa-tdnn: Emphasized channel attention, propagation and aggregation in tdnn based speaker verification,'' in \emph{Proc. Interspeech}, 2020, pp. 3830--3834.

\bibitem{ecapa++}
B.~Liu and Y.~Qian, ``Ecapa++: Fine-grained deep embedding learning for tdnn based speaker verification,'' in \emph{Proc. Interspeech}, 2023.

\bibitem{dense-residual}
Y.~Liu, Y.~Song, I.~McLoughlin, L.~Liu, and L.~Dai, ``An effective deep embedding learning method based on dense-residual networks for speaker verification,'' in \emph{IEEE International Conference on Acoustics, Speech and Signal Processing (ICASSP)}, 2021, pp. 6668--6672.

\bibitem{adaptive-cnn}
S.-H. Kim and Y.-H. Park, ``Adaptive convolutional neural network for text-independent speaker recognition,'' in \emph{Proc. Interspeech}, 2021, pp. 66--70.

\bibitem{df-resnet}
B.~Liu, Z.~Chen, S.~Wang, H.~Wang, B.~Han, and Y.~Qian, ``Df-resnet: Boosting speaker verification performance with depth-first design,'' in \emph{Proc. Interspeech}, 2022, pp. 296--300.

\bibitem{df-resnet-journal}
B.~Liu, Z.~Chen, and Y.~Qian, ``Depth-first neural architecture with attentive feature fusion for efficient speaker verification,'' \emph{IEEE Transactions on Audio, Speech, and Language Processing}, vol.~31, pp. 1825--1838, 2023.

\bibitem{mlp}
B.~Han, Z.~Chen, B.~Liu, and Y.~Qian, ``Mlp-svnet: A multi-layer perceptrons based network for speaker verification,'' in \emph{IEEE International Conference on Acoustics, Speech and Signal Processing (ICASSP)}, 2022, pp. 7522--7526.

\bibitem{pooling1}
K.~Okabe, T.~Koshinaka, and K.~Shinoda, ``Attentive statistics pooling for deep speaker embedding,'' in \emph{Proc. Interspeech}, 2018, pp. 2252--2256.

\bibitem{pooling2}
Y.~Zhu, T.~Ko, D.~Snyder, B.~Mak, and D.~Povey, ``Self-attentive speaker embeddings for text-independent speaker verification,'' in \emph{Proc. Interspeech}, 2018, pp. 3573--3577.

\bibitem{pooling3}
M.~India, P.~Safari, and J.~Hernando, ``Self multi-head attention for speaker recognition,'' in \emph{Proc. Interspeech}, 2019, pp. 4305--4309.

\bibitem{pooling4}
Z.~Gao, Y.~Song, I.~McLoughlin, P.~Li, Y.~Jiang, and L.~Dai, ``Improving aggregation and loss function for better embedding learning in end-to-end speaker verification system,'' in \emph{Proc. Interspeech}, 2019, pp. 361--365.

\bibitem{pooling5}
S.~Wang, Y.~Yang, Y.~Qian, and K.~Yu, ``Revisiting the statistics pooling layer in deep speaker embedding learning,'' in \emph{International Symposium on Chinese Spoken Language Processing (ISCSLP)}, 2021, pp. 1--5.

\bibitem{sv-loss1}
C.~Zhang and K.~Koishida, ``End-to-end text-independent speaker verification with triplet loss on short utterances.'' in \emph{Proc. Interspeech}, 2017, pp. 1487--1491.

\bibitem{sv-loss2}
Y.~Liu, L.~He, and J.~Liu, ``Large margin softmax loss for speaker verification,'' in \emph{Proc. Interspeech}, 2019, pp. 2873--2877.

\bibitem{sv-loss3}
X.~Xiang, S.~Wang, H.~Huang, Y.~Qian, and K.~Yu, ``Margin matters: Towards more discriminative deep neural network embeddings for speaker recognition,'' \emph{arXiv preprint arXiv:1906.07317}, 2019.

\bibitem{sv-loss4}
L.~Li, R.~Nai, and D.~Wang, ``Real additive margin softmax for speaker verification,'' in \emph{IEEE International Conference on Acoustics, Speech and Signal Processing (ICASSP)}, 2022, pp. 7527--7531.

\bibitem{ResNet}
K.~He, X.~Zhang, S.~Ren, and J.~Sun, ``Deep residual learning for image recognition,'' in \emph{Proceedings of the IEEE conference on computer vision and pattern recognition}, 2016, pp. 770--778.

\bibitem{duality-att}
L.~Zhang, Q.~Wang, and L.~Xie, ``Duality temporal-channel-frequency attention enhanced speaker representation learning,'' in \emph{IEEE Automatic Speech Recognition and Understanding Workshop (ASRU)}, 2021, pp. 206--213.

\bibitem{dpnet}
B.~Liu, Z.~Chen, and Y.~Qian, ``Dual path embedding learning for speaker verification with triplet attention,'' in \emph{Proc. Interspeech}, 2022, pp. 291--295.

\bibitem{simple-att}
X.~Qin, N.~Li, C.~Weng, D.~Su, and M.~Li, ``Simple attention module based speaker verification with iterative noisy label detection,'' in \emph{IEEE International Conference on Acoustics, Speech and Signal Processing (ICASSP)}, 2022, pp. 6722--6726.

\bibitem{aff}
B.~Liu, Z.~Chen, and Y.~Qian, ``Attentive feature fusion for robust speaker verification,'' in \emph{Proc. Interspeech}, 2022, pp. 286--290.

\bibitem{cnsrc-2022}
Z.~Chen, B.~Liu, B.~Han, L.~Zhang, and Y.~Qian, ``The sjtu x-lance lab system for cnsrc 2022,'' \emph{arXiv preprint arXiv:2206.11699}, 2022.

\bibitem{kd-shuai}
W.~Shuai, Y.~Yexin, W.~Tianze, Q.~Yanmin, and Y.~Kai, ``Knowledge distillation for small foot-print deep speaker embedding,'' in \emph{IEEE International Conference on Acoustics, Speech and Signal Processing (ICASSP)}, 2019, pp. 6021--6025.

\bibitem{skd}
B.~Liu, H.~Wang, Z.~Chen, S.~Wang, and Y.~Qian, ``Self-knowledge distillation via feature enhancement for speaker verification,'' in \emph{IEEE International Conference on Acoustics, Speech and Signal Processing (ICASSP)}, 2022, pp. 7542--7546.

\bibitem{binary-lm}
T.~Zhu, X.~Qin, and M.~Li, ``Binary neural network for speaker verification,'' in \emph{Proc. Interspeech}, 2021, pp. 86--90.

\bibitem{quant-ljy}
J.~Li, W.~Liu, Z.~Zhang, J.~Wang, and T.~Lee, ``Model compression for dnn-based text-independent speaker verification using weight quantization,'' in \emph{Proc. Interspeech}, 2023, pp. 1988--1992.

\bibitem{julien}
J.~Balian, R.~Tavarone, M.~Poumeyrol, and A.~Coucke, ``Small footprint text-independent speaker verification for embedded systems,'' in \emph{IEEE International Conference on Acoustics, Speech and Signal Processing (ICASSP)}, 2021, pp. 6164--6168.

\bibitem{ecapa-lite}
Q.~Lin, L.~Yang, X.~Wang, X.~Qin, J.~Wang, and M.~Li, ``Towards lightweight applications: Asymmetric enroll-verify structure for speaker verification,'' in \emph{IEEE International Conference on Acoustics, Speech and Signal Processing (ICASSP)}, 2022, pp. 7067--7071.

\bibitem{cs-ctcsconv1d}
L.~Cai, Y.~Yang, X.~Chen, W.~Tu, and H.~Chen, ``Cs-ctcsconv1d: Small footprint speaker verification with channel split time-channel-time separable 1-dimensional convolution,'' in \emph{Proc. Interspeech}, 2022, pp. 326--330.

\bibitem{kd}
G.~Hinton, O.~Vinyals, and J.~Dean, ``Distilling the knowledge in a neural network,'' \emph{arXiv preprint arXiv:1503.02531}, 2015.

\bibitem{deep-compress}
S.~Han, H.~Mao, and W.~J.~Dally, ``Deep compression: Compressing deep neural networks with pruning, trained quantization and huffman coding,'' in \emph{International Conference on Learning Representations}, 2016.

\bibitem{apot}
Y.~Li, X.~Dong, and W.~Wang, ``Additive powers-of-two quantization: An efficient non-uniform discretization for neural networks,'' in \emph{International Conference on Learning Representations}, 2020.

\bibitem{bnn}
M.~Courbariaux, I.~Hubara, D.~Soudry, R.~El-Yaniv, and Y.~Bengio, ``Binarized neural networks: Training deep neural networks with weights and activations constrained to +1 or -1,'' \emph{arXiv preprint arXiv::1602.02830}, 2016.

\bibitem{siman}
M.~Lin, R.~Ji, Z.~Xu, B.~Zhang, F.~Chao, C.-W. Lin, and L.~Shao, ``Siman: Sign-to-magnitude network binarization,'' \emph{IEEE Transactions on Pattern Analysis and Machine Intelligence}, vol.~45, no.~5, pp. 6277--6288, 2023.

\bibitem{adaptive-quant}
H.~Wang, B.~Liu, and Y.~Qian, ``Adaptive neural network quantization for lightweight speaker verification,'' in \emph{Proc. Interspeech}, 2023, pp. 5331--5335.

\bibitem{kd-zly}
L.~Zhang, Z.~Chen, and Y.~Qian, ``Knowledge distillation from multi-modality to single-modality for person verification,'' in \emph{Proc. Interspeech}, 2021, pp. 1897--1901.

\bibitem{thin-resnet2}
W.~Cai, J.~Chen, and M.~Li, ``Exploring the encoding layer and loss function in end-to-end speaker and language recognition system,'' in \emph{Proc. Odyssey}, 2018, pp. 74--81.

\bibitem{fast-resnet}
J.~S. Chung, J.~Huh, S.~Mun, M.~Lee, H.-S. Heo, S.~Choe, C.~Ham, S.~Jung, B.-J. Lee, and I.~Han, ``In defence of metric learning for speaker recognition,'' in \emph{Proc. Interspeech}, 2020, pp. 2977--2981.

\bibitem{uniform}
I.~Hubara, M.~Courbariaux, D.~Soudry, R.~El-Yaniv, and Y.~Bengio, ``Quantized neural networks: Training neural networks with low precision weights and activations,'' \emph{Journal of Machine Learning Research}, vol.~18, pp. 1--30, 2018.

\bibitem{pot}
A.~Zhou, A.~Yao, Y.~Guo, L.~Xu, and Y.~Chen, ``Incremental network quantization: Towards lossless cnns with low-precision weights,'' in \emph{International Conference on Learning Representations}, 2017.

\bibitem{ada_round}
M.~Nagel, R.~A. Amjad, M.~Van~Baalen, C.~Louizos, and T.~Blankevoort, ``Up or down? adaptive rounding for post-training quantization,'' in \emph{Proceedings of the 37th International Conference on Machine Learning}, 2020, pp. 7197--7206.

\bibitem{brecq}
Y.~Li, R.~Gong, X.~Tan, Y.~Yang, P.~Hu, Q.~Zhang, F.~Yu, W.~Wang, and S.~Gu, ``Brecq: Pushing the limit of post-training quantization by block reconstruction,'' in \emph{International Conference on Learning Representations}, 2021.

\bibitem{dorefa}
S.~Zhou, Y.~Wu, Z.~Ni, X.~Zhou, H.~Wen, and Y.~Zou, ``Dorefa-net: Training low bitwidth convolutional neural networks with low bitwidth gradients,'' \emph{arXiv preprint arXiv:1606.06160}, 2016.

\bibitem{pact}
J.~Choi, Z.~Wang, S.~Venkataramani, P.~I.-J. Chuang, V.~Srinivasan, and K.~Gopalakrishnan, ``Pact: Parameterized clipping activation for quantized neural networks,'' \emph{arXiv preprint arXiv:1805.06085}, 2018.

\bibitem{fake-quant}
B.~Jacob, S.~Kligys, B.~Chen, M.~Zhu, M.~Tang, A.~Howard, H.~Adam, and D.~Kalenichenko, ``Quantization and training of neural networks for efficient integer-arithmetic-only inference,'' in \emph{Proceedings of the IEEE conference on computer vision and pattern recognition}, 2018, pp. 2704--2713.

\bibitem{ste}
Y.~Bengio, N.~L\'eonard, and A.~Courville, ``Estimating or propagating gradients through stochastic neurons for conditional computation,'' \emph{arXiv preprint arXiv:1308.3432}, 2013.

\bibitem{hawq}
Z.~Dong, Z.~Yao, A.~Gholami, M.~Mahoney, and K.~Keutzer, ``Hawq: Hessian aware quantization of neural networks with mixed-precision,'' in \emph{IEEE/CVF International Conference on Computer Vision (ICCV)}, 2019, pp. 293--302.

\bibitem{hawq-v2}
Z.~Dong, Z.~Yao, D.~Arfeen, A.~Gholami, M.~W. Mahoney, and K.~Keutzer, ``Hawq-v2: Hessian aware trace-weighted quantization of neural networks,'' in \emph{Advances in Neural Information Processing Systems}, 2020, pp. 18\,518--18\,529.

\bibitem{k-means}
M.~I. Abiodun, E.~E. Absalom, A.~Laith, A.~Belal, and H.~Jia, ``K-means clustering algorithms: A comprehensive review, variants analysis, and advances in the era of big data,'' \emph{Information Sciences}, vol. 622, pp. 178--210, 2023.

\bibitem{inq}
A.~Zhou, A.~Yao, Y.~Guo, L.~Xu, and Y.~Chen, ``Incremental network quantization: Towards lossless cnns with low-precision weights,'' in \emph{International Conference on Learning Representations}, 2017.

\bibitem{stoch_quant}
D.~Yinpeng, J.~Li, and R.~Ni, ``Learning accurate low-bit deep neural networks with stochastic quantization,'' in \emph{Proceedings of the British Machine Vision Conference (BMVC)}, 2017, pp. 189.1--189.12.

\bibitem{n2uq}
Z.~Liu, K.-T. Cheng, D.~Huang, E.~Xing, and Z.~Shen, ``Nonuniform-to-uniform quantization: Towards accurate quantization via generalized straight-through estimation,'' in \emph{Proceedings of the IEEE conference on computer vision and pattern recognition}, 2022, pp. 4942--4952.

\bibitem{voxceleb1}
A.~Nagrani, J.~S. Chung, and A.~Zisserman, ``Voxceleb: A large-scale speaker identification dataset,'' in \emph{Proc. Interspeech}, 2017, pp. 2616--2620.

\bibitem{voxceleb2}
J.~S. Chung, A.~Nagrani, and A.~Zisserman, ``Voxceleb2: Deep speaker recognition,'' in \emph{Proc. Interspeech}, 2018, pp. 1086--1090.

\bibitem{speed_perturb}
W.~Wang, D.~Cai, X.~Qin, and M.~Li, ``The dku-dukeece systems for voxceleb speaker recognition challenge 2020,'' \emph{arXiv preprint arXiv:2010.12731}, 2020.

\bibitem{online_data_aug}
W.~Cai, J.~Chen, J.~Zhang, and M.~Li, ``On-the-fly data loader and utterance-level aggregation for speaker and language recognition,'' \emph{IEEE Transactions on Audio, Speech, and Language Processing}, vol.~28, pp. 1038--1051, 2020.

\bibitem{musan}
D.~Snyder, G.~Chen, and D.~Povey, ``Musan: a music, speech, and noise corpus,'' \emph{arXiv preprint arXiv:1510.08484}, 2015.

\bibitem{rir}
T.~Ko, V.~Peddinti, D.~Povey, M.~L. Seltzer, and S.~Khudanpur, ``A study on data augmentation of reverberant speech for robust speech recognition,'' in \emph{IEEE International Conference on Acoustics, Speech and Signal Processing (ICASSP)}, 2017, pp. 5220--5224.

\bibitem{specaug}
D.~S. Park, W.~Chan, Y.~Zhang, C.-C. Chiu, B.~Zoph, E.~D. Cubuk, and Q.~V. Le, ``Specaugment: A simple data augmentation method for automatic speech recognition,'' in \emph{Proc. Interspeech}, 2019, pp. 2613--2617.

\bibitem{adamw}
I.~Loshchilov and F.~Hutter, ``Decoupled weight decay regularization,'' in \emph{International Conference on Learning Representations}, 2019.

\bibitem{asnorm}
Z.~N. Karam, W.~M. Campbell, and N.~Dehak, ``Towards reduced false-alarms using cohorts,'' in \emph{IEEE International Conference on Acoustics, Speech and Signal Processing (ICASSP)}, 2011, pp. 4512--4515.

\bibitem{admm}
J.~Xu, J.~Yu, S.~Hu, X.~Liu, and H.~Meng, ``Mixed precision low-bit quantization of neural network language models for speech recognition,'' \emph{IEEE Transactions on Audio, Speech, and Language Processing}, vol.~29, pp. 3679--3693, 2021.

\bibitem{twn}
B.~Liu, F.~Li, X.~Wang, B.~Zhang, and J.~Yan, ``Ternary weight networks,'' in \emph{IEEE International Conference on Acoustics, Speech and Signal Processing (ICASSP)}, 2023, pp. 1--5.

\end{thebibliography}


 





\end{document}